\documentclass{issue-lpr}
\usepackage[english]{babel}
\usepackage{times}
\usepackage{psfrag}
\usepackage{amsmath}
\usepackage{amsfonts}
\usepackage{amssymb}
\usepackage{amstext}
\newcommand{\beq}{\begin{equation}}
\newcommand{\eeq}{\end{equation}}
\newcommand{\beqa}{\begin{eqnarray}}
\newcommand{\eeqa}{\end{eqnarray}}
\newcommand{\beqar}{\begin{eqnarray*}}
\newcommand{\eeqar}{\end{eqnarray*}}

\newcommand{\iu}{\text{i}}
\newcommand{\ee}{\text{e}}

\newcommand{\proj}[1]{\ket{#1}\bra{#1}}
\newcommand{\bra}[1]{\langle #1 |}
\newcommand{\ket}[1]{| #1 \rangle}
\newcommand{\Bra}{\langle}
\newcommand{\Ket}{\rangle}

\definecolor{mygray}{gray}{0.94}
\definecolor{myyellow}{rgb}{1,1,0.6}
\definecolor{untersch1}{rgb}{0,0.6,0}
\definecolor{untersch2}{rgb}{0.8,0.1,0.2}
\definecolor{untersch3}{rgb}{0,0.7,0.3}
\definecolor{untersch4}{rgb}{0.4,0,0.6}
\definecolor{highlight}{rgb}{1,0,0}
\definecolor{important}{rgb}{1,0,0}

\definecolor{lightblue}{rgb}{0.75,0.8,1}
\definecolor{nblue}{rgb}{0,0,1}
\definecolor{ngreen}{rgb}{0,1,0}
\definecolor{graublau}{rgb}{0.85,0.85,0.9}
\definecolor{gruen}{rgb}{0.0,0.4,0.0}
\definecolor{green4}{rgb}{0.0,0.5,0.0}
\definecolor{alexblue}{rgb}{0.8,0.64,0.0}
\definecolor{strangeyellow}{rgb}{0.875,1.0,0.08}
\definecolor{verylightgray}{gray}{0.99}
\definecolor{meinrot}{rgb}{0.8,0.1,0.2}
\definecolor{lightocker}{rgb}{1,0.9,0.6}
\definecolor{orange}{rgb}{1,0.5,0}
\definecolor{darkorange}{rgb}{0.6,0.1,0}
\definecolor{rotorange}{rgb}{0.6,0.1,0}
\definecolor{rotorange2}{rgb}{0.4,0.2,0.2}
\definecolor{brown4}{rgb}{0.6,0.2,0}
\definecolor{brown2}{rgb}{0.7,0.4,0}
\definecolor{red3}{rgb}{0.7,0,0}
\definecolor{red2}{rgb}{0.8,0,0}

\graphicspath{{./}{figs/}}

\setpages{1}

\setvolume[1]{1}

\setyear{2008}%

\setdoi{20070001}%

\begin{document}

\titlefigure[clip,width=\linewidth,height=46mm]{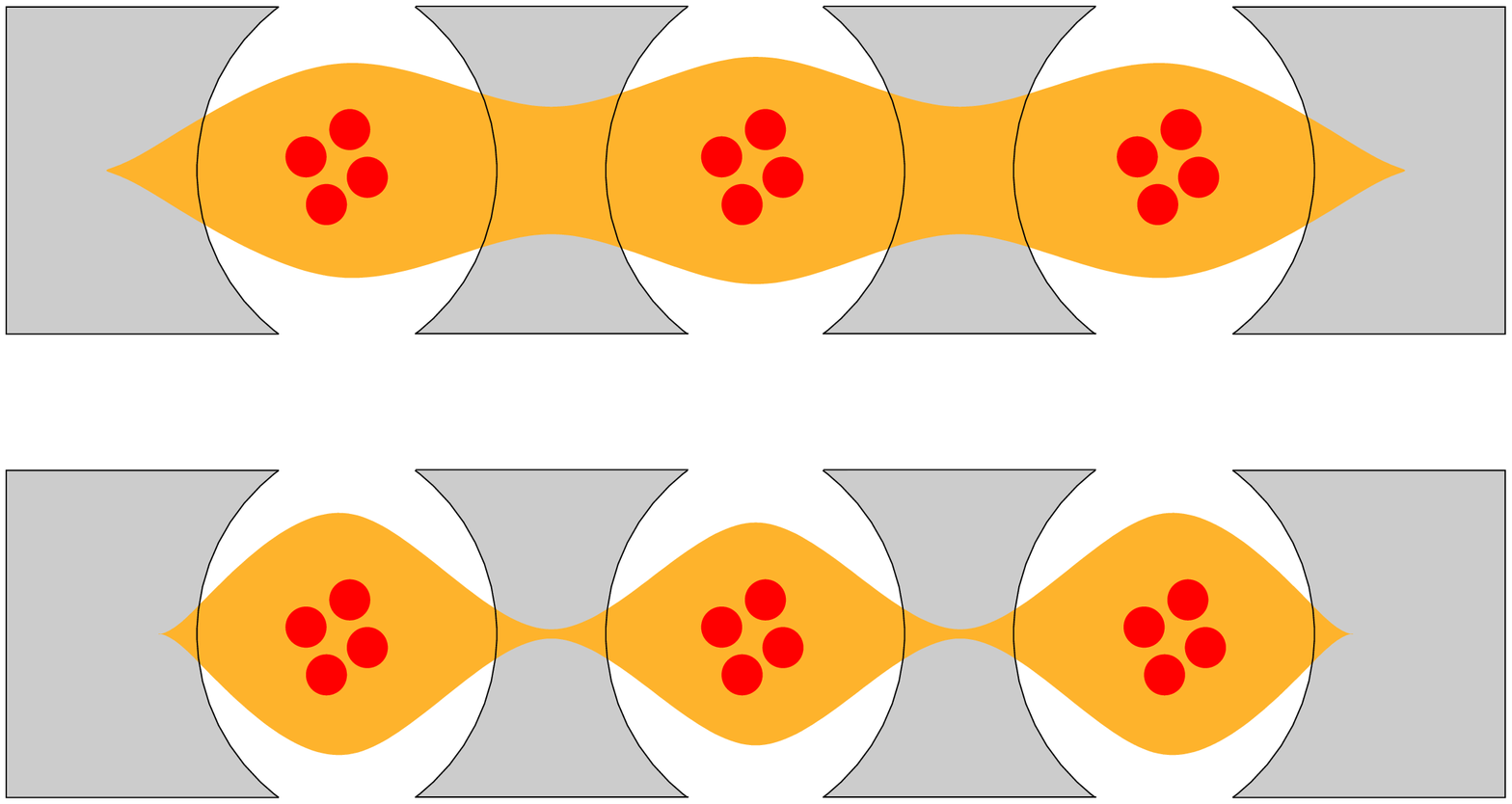}

\titlefigurecaption{{\bf Crystal light:} Photons (yellow) that ordinarily move freely through an array of cavities (top) might be frozen in place by their mutual repulsion (bottom). Such repulsion can be induced by laser pumping of suitable atoms (red) placed in the cavities.}

\abstract{The increasing level of experimental control over atomic and optical systems gained in the past years have paved the way for the exploration of new physical regimes in quantum optics and atomic physics, characterised by the appearance of quantum many-body phenomena, originally encountered only in condensed-matter physics, and the possibility of experimentally accessing them in a more controlled manner. In this review article we survey recent theoretical studies concerning the use of cavity quantum electrodynamics to create quantum many-body systems. Based on recent experimental progress in the fabrication of arrays of interacting micro-cavities and on their coupling to atomic-like structures in several different physical architectures, we review proposals on the realisation of paradigmatic many-body models in such systems, such as the Bose-Hubbard and the anisotropic Heisenberg models. Such arrays of coupled cavities offer interesting properties as simulators of quantum many-body physics, including the full addressability of individual sites and the accessibility of inhomogeneous models.}

\title{Quantum Many-Body Phenomena in Coupled Cavity Arrays}

\author{Michael J. Hartmann\inst{1,2,3,*}, Fernando G.S.L. Brand\~ao\inst{2,3} and Martin B. Plenio\inst{2,3}}

\institute{%
 Technische Universit\"at M\"unchen, Physik Department I,James Franck Str., 85748 Garching, Germany
\and
  Institute for Mathematical Sciences, Imperial College London,
SW7 2PG, United Kingdom
\and
  QOLS, The Blackett Laboratory, Imperial College London, Prince Consort Road,
SW7 2BW, United Kingdom
}

\mail{e-mail: michael.hartmann@ph.tum.de}

\titlerunning{Quantum Many-Body Phenomena in Coupled Cavity Arrays}
\authorrunning{Michael J. Hartmann et. al.}

\keywords{quantum many-body models, polaritons, cavity QED, photon blockade}
\pacs{42.50.Dv,42.50.Pq,03.67.Bg,75.10.Jm} 

\received{\ldots}
\published{\ldots}
\maketitle
\markboth{thedoi}{thedoi}
\tableofcontents

\section{Introduction}

Quantum optics and atomic physics are concerned with the interaction of light and matter in the quantum regime, typically on the scale of a few atoms and light quanta. As atoms do not naturally interact strongly with each other and with light, the physics of such systems can usually be very well understood by ignoring collective effects and treating the interactions perturbatively. This is a much simpler and cleaner situation than the one encountered in the condensed matter context. There, strong interactions among the basic constituents, such as nuclei and electrons, lead to the emergence of completely new physics when one considers a mesoscopic or macroscopic number of interacting particles. Thus, even though the fundamental interactions between the constituent particles are usually known, it is challenging to fully describe the properties of such systems.

The higher level of isolation of quantum optical systems have permitted, over the past decades, the development of experimental techniques for their manipulation with an unprecedented level of control. Highlights include the invention of trapping and cooling techniques \cite{MS99}, which have found applications e.g. in spectroscopy and precision measurements and which culminated in the realisation of a Bose-Einstein condensate (BEC) of neutral atoms \cite{Leg01}; the manipulation of the state of individual atomic and photonic systems at the quantum level \cite{LBM+03,RBH01} and its use in the study of the foundations of quantum theory \cite{RBH01,Zei99}; and more recently the realisation of primitives for the processing of quantum information and for the implementation of quantum computation in several different quantum atomic and optical set-ups \cite{MM02}.  

This progress raised the possibility of engineering strong interactions in these systems and of realising such strongly correlated models in large scales. This is of interest from a fundamental point of view, as in this way we can form new phases of matter for atoms, photons, or even molecules that do not exist outside the laboratory. Moreover, due to the high level of experimental control over them, it is possible to study many-body physics in a much cleaner manner than it is possible employing condensed-matter systems, in which the strong interaction with the environment and the very short space- and time-scales involved inhibit a precise control over their static and dynamic properties. The use of one quantum mechanical system to simulate the physics of a different one is an idea that dates back to Feynman \cite{Fey82} and has attracted considerable interest recently in the field of quantum information science \cite{MM02}.

A very successful example in this direction is the study of cold trapped atoms in optical lattices \cite{BDZ08}. By loading a BEC of neutral atoms in a periodic optical lattice, formed by the off-resonant dipole interaction of atoms with overlapping standing waves of counter propagating lasers, the otherwise weak interactions of the atoms can be increased to such an extend that a strongly correlated many-body model is formed \cite{JBC+98}. This idea led to several important experiments, such as the realisation of the Bose-Hubbard model \cite{GME+02} and of a Tonks-Girardeau gas \cite{Par04} and is currently an active area of theoretical and experimental research \cite{BDZ08,Lew07}.

Also of note is the early study of strongly correlated many-body models in Josephson junctions arrays \cite{MSS01}. Although a Josephson junction is a mesoscopic solid state device, it behaves in many respects just like the type of systems studied in quantum optics, as the relevant physics can be understood by looking at a few quantum levels. In this context, arrays of interacting Josephson junctions have been used to reproduce properties of bosonic particles. For example, the Mott insulator phase \cite{Zan92} and the superconductor-to-Mott insulator transition \cite{OM96} have been observed in Josephson junctions arrays.

Finally, theoretical proposals for using trapped ions for realising many-body models have been put forwarded \cite{MW01,PC04a,PC04b,RPTS08} and a first benchmark experiment in this direction has been reported recently \cite{Fri08}.

This article is devoted to review theoretical proposals for the use of the atom-light interaction in coupled micro-cavity arrays to create strongly correlated many-body models.
This research was initiated in \cite{HBP06} and shortly after in \cite{ASB07,GTCH06} (see also \cite{Ill06}). Since then a substantial number of publications have extended these ideas in a variety of directions. These include photonic Mott insulators \cite{HP07,HBP07b}, spin models \cite{HBP07a,JXL07,KA08,LGG07,CAB08,CAB08b}, dynamical effects \cite{HP08,OIK08,ZLS07,HZSS07,ZGSS08}, multi component models \cite{HBP07c}, phase diagrams \cite{RF07,AHTL08,IOK08,MCT+08,ZSU08}, alternative physical realisations \cite{Na08,CGM+07,PAK07,LL08,XGZ07}, experimental signatures \cite{RFS08,HLSS08} and quantum information applications \cite{ASYE07,AK08,BAB07,AMB07,CAB07,NY08}. In fact, historically, some quantum information proposals \cite{ASYE07} were presented before the theory for effective many-body models \cite{HBP06,ASB07,GTCH06}.

Recent experimental progress in the fabrication of micro-cavity arrays and the realisation of the quantum regime in the interaction of atomic-like structures and quantised electromagnetic modes inside those \cite{AKS+03,ADW+06,BSP+06,AAS+03,Hen07,Wal04,TGD+07,CSD+07} opened up the possibility of using them as quantum simulators of many-body physics. The first motivation for this study is the desire to make photons or combined photonic-atomic excitations enter new quantum states, which do not naturally appear in nature. The second is that this set-up offer advantages over other proposals for the realisation of strongly interacting many-body models in quantum optical systems. Due to the small separations between neighbouring sites, it is experimentally very challenging to access individual sites in Josephson junctions arrays and optical lattices. Neighbouring sites in coupled cavity arrays, on the other hand, are usually separated by dozens of micrometres and can therefore be accurately accessed by optical frequencies. This allows e.g. the measurement of local properties as well as the exploration of inhomogeneous systems.

As the field of coupled cavity arrays is still quite young and many preprints are currently being reviewed, we have restricted ourselves here to only present details of work that has been published in peer reviewed journals before completion of the present text. This has been done in order to avoid any interferences with the policies of journals that consider the work in question for publication.

The organisation of the paper is as follows: In section \ref{cQED} we discuss the use of cavities to enhance the interaction of light and matter and in section \ref{cca} we develop the basic description for an array of interacting cavities. The following section \ref{BHmodel} reviews the theory for engineering effective Bose-Hubbard Hamiltonians with polaritons and photons. In section \ref{sec:JC} we discuss a second class of effective Hamiltonians that exhibits a phase diagram similar to the Bose-Hubbard model: Jaynes-Cummings based many body models. The phase diagrams of both classes are then reviewed in section \ref{sec:ph_diag}. Effective spin lattice Hamiltonians can also be generated in coupled cavity arrays and we review these approaches in section \ref{sec:spin}.
In section \ref{sec:impl} we then discuss several experimental platforms which appear to be suitable for a future implementation of the theory approach presented in this review.
We finish with a summary in section \ref{conc}.


\section{Cavity QED} \label{cQED}

Usually photons interact only over a very short time, and hence very weakly, with atoms. This interaction can be substantially increased if the atoms and photons are trapped inside a resonator. In the simplest situation, we can consider the coupling of a single light mode with a single two level system inside a cavity. Using the dipole and the rotating wave approximations their dynamics is governed by the Jaynes-Cummings Hamiltonian,
\begin{equation} 
H = \hbar \omega_C a^{\cal y}a + \hbar \omega_0 \ket{e}\bra{e} + \hbar g(a^{\cal y}\ket{g}\bra{e} + a\ket{e}\bra{g}),
\end{equation}
where $\omega_C, \omega_0$ are the frequencies of the resonant mode of the cavity and of the atomic transition, respectively, $g$ is Jaynes-Cummings coupling between the cavity mode and the two level system, $a^{\cal y}$ is the creation operator of a photon in the resonant cavity mode, and $\ket{g}, \ket{e}$ are the ground and excited states of the two level system. The Rabi frequency $g$ is given by
\begin{equation}
g = \vec{d} f(r_0) \sqrt{\frac{\omega_C}{2  \epsilon_0 V_{\text{mode}}}}\, ,
\end{equation}
where $V_{\text{mode}}$ is the mode volume of the cavity, $f(r_0)$ is the mode-function at the position $r_0$ of the two level system, $\vec{d}$ is the electric dipole transition moment, and $\epsilon_0$ is the free-space permittivity.

\begin{figure}
\centering
\psfrag{g4}{\hspace{-0.04cm}\color{red2}$\gamma$\normalcolor}
\psfrag{g3}{\hspace{-0.1cm}\color{red2}$\gamma$\normalcolor}
\psfrag{k}{\hspace{-0.04cm}\color{green4}$\kappa$\normalcolor}
\psfrag{g13}{\hspace{-0.04cm}\color{green4}$g$\normalcolor}
\psfrag{g24}{\hspace{-0.1cm}\color{green4}$g$\normalcolor}
\psfrag{o}{\hspace{0.02cm}\color{nblue}$\Omega$\normalcolor}
\psfrag{d2}{\hspace{-0.16cm}\raisebox{-0.06cm}{$\Delta$}}
\psfrag{d1}{\hspace{-0.02cm}\raisebox{-0.06cm}{$\delta$}}
\psfrag{w1}{\raisebox{-0.04cm}{\hspace{-0.05cm}$\omega$}}
\psfrag{w2}{\raisebox{-0.04cm}{\hspace{-0.05cm}$\omega$}}
\psfrag{1}{\hspace{-0.06cm}\raisebox{-0.18cm}{$1$}}
\psfrag{2}{\hspace{-0.06cm}\raisebox{-0.18cm}{$2$}}
\psfrag{3}{\hspace{-0.06cm}\raisebox{-0.0cm}{$3$}}
\psfrag{4}{\hspace{-0.08cm}\raisebox{-0.0cm}{$4$}}
\includegraphics[width=.7\linewidth]{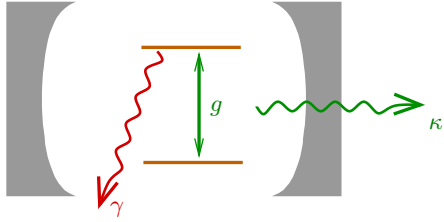}
\caption{\label{cavityqedfig} Cavity QED: A two level atom interacts via a dipole coupling $g$ with the photons in the cavity. Excitations a lost via spontaneous emission at a rate $\gamma$ and cavity decay at a rate $\kappa$.}
\end{figure}

In addition to the coherent interaction, there are two main loss processes that affect the dynamics of the system: spontaneous emission from level $\ket{e}$ to level $\ket{g}$ at a rate $\gamma$ and leaking out of photons of the cavity mode at rate $\kappa$, see figure \ref{cavityqedfig}. The cavity decay rate $\kappa$ is connected to the quality factor of the cavity $Q$ by $Q = \omega_C/(2\kappa)$. Although other processes, such as thermal motion of the atom or dephasing due to background electromagnetic fields, also contribute to the losses of the system, their effect are usually much smaller and can be disregarded in many applications.

The strong coupling regime of cavity QED is reached when the cooperativity factor, given by the the vacuum Rabi frequency to the square over the product of the spontaneous emission and cavity decay rates,
\begin{equation} \label{cooperativity}
 \xi := \frac{g^2}{2 \kappa \gamma} \, ,
\end{equation}
is much larger than one: $\xi \gg 1$. In this regime the coherent part of the evolution dominates over the decoherence processes and quantum dynamics of the joint atom cavity mode system can be observed. Such a regime has first been achieved in seminal experiments for microwave \cite{RBH01} and later optical frequencies \cite{TRK92}, both using a single atom inside a Fabry-P\'erot cavity formed between two miniature spherical mirrors. In these works, the strong coupling regime has been employed e.g. to create conditional quantum dynamics and entanglement between the photonic and atomic degrees of freedom \cite{RBH01,THL+95} and to perform quantum non-demolition measurements \cite{RBH01}. Recently a strong coupling regime with very high cooperativity factors has also been achieved with artificial atoms formed by superconducting structures that interact with the electromagnetic modes of stripline resonator \cite{Wal04}.

Most importantly for our purposes the strong coupling regime led to a realisation of the photon blockade effect \cite{Bir05}, where, due to the strong interaction of the cavity mode with atoms, a single photon can modify the resonance frequency of the cavity mode in such a way that a second photon can not enter the cavity before the first leaks out.

In this review we are interested in the situation where several cavities operating in the strong coupling regime are coupled to each other. It turns out that the Fabry-P\'erot architecture is not very suitable for the coupling of separate cavities. Nonetheless, several new cavity QED set-ups have recently emerged in which (i) the strong coupling regime has been achieved and (ii) the construction of arrays of coupled cavities has been realised, or at least seem reasonable to be so. These new systems, which we briefly review in section \ref{sec:impl}, are the strongest candidates for realising the proposals that we review in this paper.

\section{Coupled Cavity Arrays} \label{cca}
\begin{figure}
	\centering
	\includegraphics[width=\linewidth]{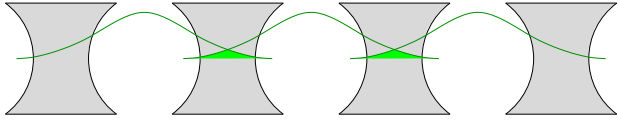}
	\caption{An array of coupled cavities. Photon hopping occurs due to the overlap (shaded green) of the light modes (green lines) of adjacent cavities.}
	\label{crystal1a}
\end{figure}
In this section we present the theoretical description for the coupling of cavities in an array. Photon hopping can occur between neighbouring cavities due to the overlap of the spacial profile of the cavity 
modes, see figure \ref{crystal1a} . In order to model such a process, we follow references \cite{HBP06,SM98,YXLS99} and describe the array of cavities by a periodic dielectric constant, $\epsilon (\vec{r}) = \epsilon (\vec{r} + R\vec{n})$, where $\vec{r}$ is a given three dimensional vector, $R$ is a constant and $\vec{n}$ labels tupels of integers. In the Coulomb gauge the electromagnetic field is represented by a vector potential $\vec{A}$ satisfying $\nabla \cdot (\epsilon(\vec{r}) \, \vec{A}) = 0$. We can expand $\vec{A}$ in Wannier functions, $\omega_{\vec{R}}$, each localised in a single cavity at location $\vec{R} = R \vec{n}$. We describe this single cavity by the dielectric function $\epsilon_{\vec{R}}(\vec{r})$. Then, from Maxwell's equations one can derive the eigenvalue equation
\begin{equation} \label{Wanniereval}
\frac{\epsilon_{\vec{R}}(\vec{r}) \, \omega_C^2}{c^2} \vec{w}_{\vec{R}} \, -
\, \nabla \times \left( \nabla \times \vec{w}_{\vec{R}} \right) = 0 \, ,
\end{equation}
where the eigenvalue $\omega_C^2$ is the square the resonance frequency of the cavity, which is independent of $\vec{R}$ due to the periodicity. Assuming that the Wannier functions decay sufficiently fast outside the cavity, only Wannier modes of nearest neighbour cavities have a non-vanishing overlap.

Introducing the creation and annihilation operators of the Wannier modes, $a_{\vec{R}}^{\dagger}$ and $a_{\vec{R}}$, the Hamiltonian of the field can be written as
\begin{eqnarray} \label{arrayham2}
\mathcal{H} &=& \omega_C \sum_{\vec{R}}
\left(  a_{\vec{R}}^{\dagger} a_{\vec{R}} + \frac{1}{2} \right) \nonumber \\
&+& 2 \omega_C \alpha \sum_{<\vec{R}, \vec{R}'>}
\left( a_{\vec{R}}^{\dagger} a_{\vec{R}'} + \text{h.c.} \right) \, .
\end{eqnarray}
Here $\sum_{< \vec{R}, \vec{R}' >}$ is the sum of all pairs of cavities which are nearest neighbours. Since $\alpha \ll 1$, we neglected counter rotating terms which contain products of two
creation or two annihilation operators of Wannier modes in deriving (\ref{arrayham2}). The coupling parameter $\alpha$ is given by \cite{YXLS99},
\begin{equation}
\alpha = \int d^3r \, \left(\epsilon_{\vec{R}}(\vec{r}) \, - \, \epsilon(\vec{r}) \right) \,
\vec{w}_{\vec{R}}^{\star} \vec{w}_{\vec{R}'} \, ; \quad 
|\vec{R} - \vec{R}'| = R \, ,
\end{equation}
and has been obtained numerically for specific models \cite{MV97}.

In deriving Hamiltonian (\ref{arrayham2}) we have assumed that all the cavities have the same resonant frequency and that the overlap of the Wannier modes $\alpha$ is the same for all cavity-cavity interactions. In practise of course there will always be some disorder in the array
and the resonance frequencies, $\omega_C (\vec{R})$, and tunnelling rates, $2 \omega_C (\vec{R}) \alpha ((\vec{R})$ will differ from cavity to cavity. The disorder in the array can even give rise to interesting effects such as the emergence of glassy phases, see \ref{sec:ph_diag}. 


\section{The Bose-Hubbard Model in Coupled Cavity Arrays} \label{BHmodel}

Although realising Hamiltonian (\ref{arrayham2}) might be interesting for quantum information propagation \cite{HRP06,PHE04}, it is not the 
type of model that one would like to quantum simulate. Indeed, the model is harmonic and can be very easily solved in terms of Bloch waves, which allows for a simple understanding of all its basic properties. The situation changes dramatically if we add an on-site interaction term. The interplay of tunnelling and interaction leads to interesting many-body physics. A paradigmatic model for such a situation is the Bose-Hubbard Hamiltonian, which reads
\begin{eqnarray} \label{bosehubbard}
H_{BH} = \mu \sum_{k} b^{\cal y}_k b_k &-& J \sum_{<k, k'>} (b_{k}^{\cal y}b_{k'} + h.c.) \nonumber \\ &+& U \sum_{k} b^{\cal y}_k b_k(b^{\cal y}_k b_k - 1),
\end{eqnarray}
where $b_k^{\cal y}$ creates a boson at site $k$, $J$ is the hopping rate, $U$ the on-site interaction strength, and $\mu$ the chemical potential. This model, which
was first discussed by Fisher \textit{et al} in Ref. \cite{Fis89,Fis90} as the bosonic counterpart of the Hubbard model \cite{Hub63},
contains two different phases at zero temperature that are separated by a quantum phase transition where the ration $U/J$ crosses a critical value. Let us briefly describe the two phases qualitatively. 

When $J \gg U$, the tunnelling term dominates and the lowest energy state of the system is a condensate of delocalised bosons: the system is in a superfluid phase. If we start to increase a repulsive on-site interaction, adding more than one particle to one site requires energy that grows quadratically with the particle number. It will thus be energetically favourable for the bosons to evenly distribute across all lattice sites and avoid moving around in order to minimise the repulsion. Above a specific value of $U/J$, the system ceases to be a superfluid and becomes a Mott insulator, with a well defined number of localised particles in each site\footnote{In the case of incommensurate filling factors, i.e. if the number of particles is not a multiple of the number of lattice sites, the system will always be in a superfluid phase.}. The superfluid-to-Mott insulator phase transition is of a quantum character and is driven by quantum fluctuations, as it occurs even at zero temperature. An order parameter for such a transition is the variance of the number of bosons in a given site, $\langle (b^{\cal y}_k b_k - \langle b^{\cal y}_k b_k \rangle )^2 \rangle$. In the superfluid phase, the number of bosons in each site fluctuates and thus the variance has a non-zero value. In the Mott insulator phase, in turn, the variance is close to zero, as the bosons tend to be localised in each site.

As we saw in section \ref{cca}, the coupling between neighbouring cavities naturally leads to the tunnelling term for photons in arrays of coupled cavities. Hence, for a realisation of a Bose-Hubbard model a way to effectively generate an interaction term
is needed. Indeed, up to harmonic terms which can easily be compensated for, the Bose-Hubbard interaction term has the form of a Kerr self non-linearity $(b_k^{\cal y})^2 b_k^2$. Although such an interaction naturally appears in some media as a result of a non-zero third order electric susceptibility, its effect is negligible on the level of individual quanta, which is a reason for the difficulty of realising nonlinear optics for individual quantum systems. Using the enhanced light-matter interaction in cavity QED, it is possible to engineer much stronger nonlinearities. Intuitively the strong interactions of the light mode with atoms inside the cavity, under particular circumstances, mediates strong nonlinear interaction among the photons of the cavity mode. The strength of the nonlinearity can be increased even further if instead of considering photons as the bosonic particles of the model, one considers polaritons, joint photonic-atomic excitations. In the next section we review the proposal for realising the Bose-Hubbard Hamiltonian using such polaritons.

\subsection{Polaritonic Bose-Hubbard model} \label{sec:pbh}
\begin{figure}
	\centering
	\includegraphics[width=\linewidth]{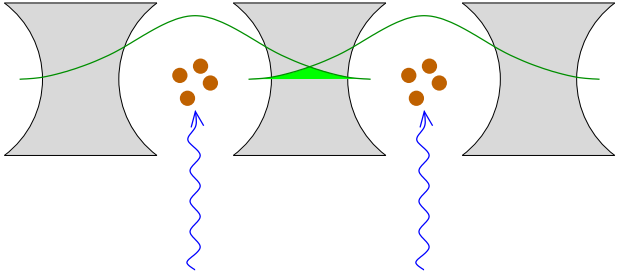}
	\caption{An array of cavities as described by our model. Photon hopping occurs due to the overlap (shaded green) of the light modes (green lines) of adjacent cavities. Atoms in each cavity (brown), which are driven by external lasers (blue) give rise to an on site potential. Reprinted with permission from \cite{HBP06}.}
	\label{crystal}
\end{figure}
\begin{figure}
	\centering
	\psfrag{g13}{\hspace{-0.04cm}$g_{13}$}
	\psfrag{g24}{\hspace{-0.16cm}$g_{24}$}
	\psfrag{o}{$\Omega$}
	\psfrag{d}{\hspace{-0.12cm}$\Delta$}
	\psfrag{d2}{\hspace{-0.04cm}$\delta$}
	\psfrag{e}{$\varepsilon$}
	\psfrag{w1}{\hspace{0.02cm}$\omega_C$}
	\psfrag{w2}{\hspace{-0.18cm}$\omega_C$}
	\psfrag{1}{\raisebox{-0.1cm}{$1$}}
	\psfrag{2}{\raisebox{-0.1cm}{$2$}}
	\psfrag{3}{$3$}
	\psfrag{4}{$4$}
	\includegraphics[width=.7\linewidth]{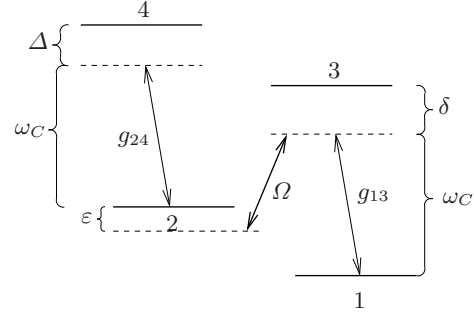}
	\caption{The level structure and the possible transitions of one atom, $\omega_C$ is the frequency of the cavity mode, $\Omega$ is the Rabi frequency of the driving by the laser, $g_{13}$ and $g_{24}$ are the parameters of the respective dipole couplings and $\delta$, $\Delta$ and $\varepsilon$ are detunings. Reprinted with permission from \cite{HBP06}.}
	\label{level}
\end{figure}

The polaritonic Bose-Hubbard model is hosted in an array of cavities that are filled with atoms of a particular 4 level structure, which
are driven with an external laser, see figures \ref{crystal} and \ref{level}. Thereby the laser drives the atoms in the same manner as in Electromagnetically Induced Transparency (EIT) \cite{FIM05}: The transitions between levels 2 and 3 are coupled to the laser field and the transitions between levels 2-4 and 1-3 couple via dipole moments to the cavity resonance mode. Levels 1 and 2 are assumed to be metastable and their spontaneous emission rates are thus negligible.

It has been shown by Imam\u{o}glu and co-workers, that this atom cavity system can exhibit a very large nonlinearity \cite{SI96,ISWD97,WI99,Gehri99}, and a similar nonlinearity has been observed experimentally \cite{Bir05}.

Considering the level structure of figure \ref{level}, in a rotating frame with respect to
$H_0 = \omega_C \left( a^{\dagger} a + \frac{1}{2} \right)$ $+$ \linebreak
$\sum_{j=1}^N \left( \omega_C \sigma_{22}^j + \omega_C \sigma_{33}^j + 2 \omega_C \sigma_{44}^j \right)$,
the Hamiltonian of the atoms in the cavity reads,
\begin{eqnarray} \label{H_manyatom}
H^I & = &
\sum_{j=1}^N \left(\varepsilon \sigma_{22}^j + \delta \sigma_{33}^j + (\Delta + \varepsilon)
\sigma_{44}^j \right) \\
& + & \, \sum_{j=1}^N \left( \Omega \, \sigma_{23}^j \, + \,
g_{13} \, \sigma_{13}^j \, a^{\dagger} \, + \,
g_{24} \, \sigma_{24}^j \, a^{\dagger} \, + \, \text{h.c.} \right)  \, , \nonumber
\end{eqnarray}
where $\sigma_{kl}^j = \ket{k_j}\bra{l_j}$ projects level $l$ of atom $j$ to level $k$ of the same atom, $\omega_C$ is the frequency of the cavity mode, $\Omega$ is the Rabi frequency of the driving by the laser and $g_{13}$ and $g_{24}$ are the parameters of the dipole coupling of the cavity mode to the respective atomic transitions.

In a cavity array, $N$ atoms in each cavity couple to the cavity mode via the interaction $H^I$ and photons tunnel between neighbouring cavities as described in equation (\ref{arrayham2}). Hence the full Hamiltonian describing this system reads
\begin{eqnarray} \label{polariton_raw}
H &=& \sum_{\vec{R}} H^I_{\vec{R}} + \omega_C \sum_{\vec{R}}
\left(  a_{\vec{R}}^{\dagger} a_{\vec{R}} + \frac{1}{2} \right) \nonumber \\
&+& 2 \omega_C \alpha \sum_{<\vec{R}, \vec{R}'>}
\left( a_{\vec{R}}^{\dagger} a_{\vec{R}'} + \text{h.c.} \right)
\end{eqnarray}
and can emulate a Bose-Hubbard model for polaritons as we will see in the following.

Assuming that all atoms interact in the same way with the cavity mode, the description can be restricted to Dicke type dressed states, in which the atomic excitations are delocalised among all the atoms. In the case where $g_{24} = 0$ and $\varepsilon = 0$ , level 4 of the atoms decouples from the dressed-state excitation manifolds \cite{WI99}. The Hamiltonian (\ref{H_manyatom}) can then be expressed in terms of the following creation (and annihilation) operators:
\begin{eqnarray} \label{polariton_operators}
p_0^{\dagger} & = & \frac{1}{B} \, \left(g S_{12}^{\dagger} - \Omega a^{\dagger} \right) , \\
p_{\pm}^{\dagger} & = & \sqrt{\frac{2}{A (A \pm \delta)}} \, \left(\Omega S_{12}^{\dagger} + g a^{\dagger} \pm
\frac{A \pm \delta}{2} S_{13}^{\dagger} \right) \nonumber ,
\end{eqnarray}
where $g = \sqrt{N} g_{13}$, $B = \sqrt{g^2 + \Omega^2}$, $A = \sqrt{4 B^2 + \delta^2}$,
$S_{12}^{\dagger} = \frac{1}{\sqrt{N}} \sum_{j=1}^N \sigma_{21}^j$
and $S_{13}^{\dagger} = \frac{1}{\sqrt{N}} \sum_{j=1}^N \sigma_{31}^j$.

The operators $p_0^{\dagger}$, $p_+^{\dagger}$ and $p_-^{\dagger}$ describe polaritons,
quasi particles formed by combinations of atom and photon excitations.
In the subspace spanned by symmetric Dicke states and in the limit of large atom numbers, $N \gg 1$, they satisfy bosonic commutation relations,
\begin{equation} \label{polariton_comm}
\left[ p_j, p_l \right] = 0 \: \: \text{and} \: \:
\left[ p_j, p_l^{\dagger} \right] = \delta_{jl} \: \: \text{for} \: \: j,l = 0,+,-.
\end{equation}
$p_0^{\dagger}$, $p_+^{\dagger}$ and $p_-^{\dagger}$ thus describe independent bosonic
particles. In terms of these polaritons, the Hamiltonian (\ref{H_manyatom}) for $g_{24} = 0$ and
$\varepsilon = 0$ reads,
\begin{equation} \label{H0polariton}
\left[H^I\right]_{g_{24} = 0, \varepsilon = 0} = 
\mu_0 \, p_0^{\dagger} p_0 + \mu_+ \, p_+^{\dagger} p_+ + \mu_- \, p_-^{\dagger} p_- \, ,
\end{equation}
where the frequencies are given by $\mu_0 = 0$, $\mu_+ = (\delta - A)/2$ and
$\mu_- = (\delta + A)/2$. 

The polaritons $p_0$ only contain atomic contributions in the two metastable states $1$ and $2$ but not in level $3$, which shows spontaneous emission. Hence, these polaritons do not populate radiating atomic levels and are thus called dark state polaritons \cite{FIM05}.

As we will discuss now, the dynamics of the dark state polaritons $p_0$ decouples from
the dynamics of the remaining species $p_+$ and $p_-$ for a suitable parameter regime.
In this regime the frequency $\mu_0$ is sufficiently separated from $\mu_+$ and $\mu_-$, such that neither the interaction with the atomic level $4$ nor the photon tunnelling between cavities can induce any mixing between the polariton species
$p_0$ and $p_{\pm}$.

\subsubsection{Polariton-polariton interactions and polariton tunnelling}

The polaritons $p_0^{\dagger}$ interact with each other via a self Kerr nonlinearity. To derive this interaction, one first needs to write the full Hamiltonian $H^I$, (\ref{H_manyatom}), in the polariton basis, expressing the operators $\sum_{j=1}^N \sigma_{22}^j$ and $a^{\dagger} \, \sum_{j=1}^N \sigma_{24}^j$ in terms of $p_0^{\dagger}$, $p_+^{\dagger}$ and $p_-^{\dagger}$. The coupling of the polaritons to the level 4 of the atoms via the dipole moment $g_{24}$ reads,
\begin{equation} \label{coupletolevel4}
g_{24} \, \left( \sum_{j=1}^N \sigma_{42}^j \, a \, + \text{h.c.} \right) \approx
- g_{24} \, \frac{g \Omega}{B^2} \, \left( S_{14}^{\dagger} \, p_0^2 \, + \text{h.c.} \right) \, ,
\end{equation}
where $S_{14}^{\dagger} = \frac{1}{\sqrt{N}} \sum_{j=1}^N \sigma_{41}^j$. In deriving (\ref{coupletolevel4}), a rotating wave approximation was used: In a frame rotating with respect to (\ref{H0polariton}), the polaritonic creation operators rotate with the frequencies $\mu_0$, $\mu_+$ and $\mu_-$. Furthermore, the operator $S_{14}^{\dagger}$ rotates at the frequency $2 \mu_0$, (c.f. (\ref{H_manyatom})). Hence, provided that 
\begin{equation} \label{rotatingwappr}
|g_{24}| \, , \, |\varepsilon| \, , \, |\Delta| \, \ll \, |\mu_+| \, , \, |\mu_-|
\end{equation}
all terms that rotate at frequencies $2 \mu_0 - (\mu_+ \pm \mu_-)$ or $\mu_0 \pm \mu_+$ or $\mu_0 \pm \mu_-$ can be neglected, which eliminates all interactions that would couple $p_0^{\dagger}$ and $S_{14}^{\dagger}$ to the remaining polariton species.

For $|g_{24 }g \Omega / B^2| \ll |\Delta|$, the coupling to level 4 can be treated perturbatively.
This results in an energy shift of $2 U$ with
\begin{equation} \label{osint}
U = - \frac{g_{24}^2}{\Delta} \,
\frac{N g_{13}^2 \, \Omega^2}{\left(N g_{13}^2 \, + \, \Omega^2 \right)^2}
\end{equation}
and in an occupation probability of the state of one $S_{14}^{\dagger}$ excitation of
$- 2 U / \Delta$, which determines an effective decay rate for the polariton $p_0^{\dagger}$ via
spontaneous emission from level 4. Note that $U > 0$ for $\Delta < 0$ and vice versa.
In a similar way, the two photon detuning $\varepsilon$ leads to an energy shift of
$\varepsilon \, g^2 \, B^{-2}$ for the polariton $p_0^{\dagger}$, which plays the role of a chemical
potential in the effective Hamiltonian.

Hence, provided (\ref{rotatingwappr}) holds, the Hamiltonian for the dark state polariton $p_0^{\dagger}$ can be written as
\begin{equation} \label{Heffect}
H_{\text{eff}} = U \, \left(p_0^{\dagger}\right)^2 \left(p_0\right)^2 \, + \, \varepsilon \, \frac{g^2}{B^2} \, p_0^{\dagger} p_0 \, ,
\end{equation}
in the rotating frame.

Let us now look at an array of interacting cavities, each under the conditions discussed above. 
The first term of the Hamiltonian (\ref{arrayham2}) has already been incorporated in the polariton analysis for one individual cavity. The second term, which describes photon tunnelling, transforms into the polariton picture via (\ref{polariton_operators}). To distinguish between the dark state polaritons in different cavities, we introduce the notation $p_{\vec{R}}^{\dagger}$ to label the polariton $p_0^{\dagger}$ in the cavity at position $\vec{R}$. The photon hopping translates to polariton hoppings according to,
\begin{eqnarray} 
a_{\vec{R}}^{\dagger} a_{\vec{R}'}
& \approx & \frac{\Omega^2}{B^2} \, p_{\vec{R}}^{\dagger} \, p_{\vec{R}'} \\
& + & \text{"terms for other polariton species"} \nonumber \, .
\end{eqnarray}
Contributions of different polaritons decouple due to the separation of their
frequencies $\mu_0$, $\mu_+$ and $\mu_-$. As a consequence the Hamiltonian for the polaritons $p_{\vec{R}}^{\dagger}$ takes on the form (\ref{bosehubbard}), with 
\begin{equation}
J = \frac{2 \omega_C \Omega^2}{N g_{13}^2 + \Omega^2} \alpha \, ,
\end{equation}
$U$ as given by equation (\ref{osint}) and $\mu = \epsilon g^2 / B^2$.

\subsubsection{Polariton lifetime: spontaneous emission and cavity decay}

We now discuss the loss sources in the system. Levels 1 and 2 are metastable and therefore have negligible decay rates on the relevant time scales. The decay mechanism for the dark state polariton then originates mainly from photons that leak out of the cavity and from the small, but non-zero, population of level 4, which leads to spontaneous emission. The resulting effective decay rate for the dark state polariton reads
\begin{equation} \label{gamma0}
\Gamma_0 = \frac{\Omega^2}{B^2}\kappa + \Theta(n - 2)\frac{g_{24}^2g^2\Omega^2}{\Delta^2 B^4}\gamma_4,
\end{equation}
where $\kappa$ is the cavity decay rate, $\gamma_4$ the spontaneous emission rate from level 4, $n$ is the average number of photons, and $\Theta$ the Heaviside step function. Assuming that $g_{13} = g_{24}$, the maximal achievable rate $U/ \Gamma_0$ can be readily seen to be $g_{13}/\sqrt{4 \kappa \Theta(n - 2)\gamma_4}$. Hence, for large cooperativity factors, $\xi \gg 1$, ratios of $U/ \Gamma_0 \gg 1$ can be achieved, which ensures that the dynamics of the Bose-Hubbard model can be observed under realistic conditions.

\subsubsection{The phase transition}

Of great interest is of course whether the quantum phase transition from a Mott insulator to a superfluid state could be observed with the present polariton approach.
In a system with on average one polariton per cavity, the Mott insulator state is characterised by the fact that the local state in each cavity is a single polariton Fock state with vanishing fluctuations of the polariton number. The superfluid phase in contrast shows polariton number fluctuations.

\begin{figure*}
\centering
\psfrag{K}{\hspace{0.12cm} \small $U$}
\psfrag{J}{\hspace{0.12cm} \small $J$}
\psfrag{data1}{\hspace{0.12cm} \small $n_1$}
\psfrag{data2}{\hspace{0.12cm} \small $F_1$}
\psfrag{t}{\hspace{-0.4cm}\raisebox{-0.3cm}{\small $t$ in $10^{-6}$ s}}
\psfrag{h}{\hspace{-0.2cm}\raisebox{0.4cm}{\small MHz}}
\psfrag{a}{\large {\bf a}}
\psfrag{b}{\large {\bf b}}
\psfrag{c}{\large {\bf c}}
\psfrag{d}{\large {\bf d}}
\psfrag{o}{\color{red} $\Omega$ \normalcolor}
\psfrag{0}{\raisebox{-0.06cm}{\tiny $0$}}
\psfrag{1000}{\raisebox{-0.06cm}{\tiny $0.1$}}
\psfrag{2000}{\raisebox{-0.06cm}{\tiny $0.2$}}
\psfrag{3000}{\raisebox{-0.06cm}{\tiny $0.3$}}
\psfrag{4000}{\raisebox{-0.06cm}{\tiny $0.4$}}
\psfrag{5000}{\raisebox{-0.06cm}{\tiny $0.5$}}
\psfrag{6000}{\raisebox{-0.06cm}{\tiny $0.6$}}
\psfrag{7000}{\raisebox{-0.06cm}{\tiny $0.7$}}
\psfrag{8000}{\raisebox{-0.04cm}{\tiny $0.8$}}
\psfrag{9000}{\raisebox{-0.06cm}{\tiny $0.9$}}
\psfrag{10000}{\raisebox{-0.04cm}{\tiny $1.0$}}
\psfrag{-0.025a}{\hspace{-0.23cm}\tiny $-0.025$}
\psfrag{-0.02a}{\hspace{-0.23cm}\tiny $-0.02$}
\psfrag{-0.015a}{\hspace{-0.23cm}\tiny $-0.015$}
\psfrag{-0.01a}{\hspace{-0.23cm}\tiny $-0.01$}
\psfrag{-0.005a}{\hspace{-0.23cm}\tiny $-0.005$}
\psfrag{0a}{\hspace{-0.1cm}\tiny $0$}
\psfrag{0.005a}{\hspace{-0.14cm}\tiny $0.005$}
\psfrag{0.01a}{\hspace{-0.14cm}\tiny $0.01$}
\psfrag{0.015a}{\hspace{-0.14cm}\tiny $0.015$}
\psfrag{0.02a}{\hspace{-0.14cm}\tiny $0.02$}
\psfrag{0.1a}{\hspace{-0.15cm}\tiny $0.1$}
\psfrag{0.2a}{\hspace{-0.15cm}\tiny $0.2$}
\psfrag{0.3a}{\hspace{-0.15cm}\tiny $0.3$}
\psfrag{0.4a}{\hspace{-0.15cm}\tiny $0.4$}
\psfrag{0.5a}{\hspace{-0.15cm}\tiny $0.5$}
\psfrag{0.6a}{\hspace{-0.15cm}\tiny $0.6$}
\psfrag{0.7a}{\hspace{-0.15cm}\tiny $0.7$}
\psfrag{0.8a}{\hspace{-0.15cm}\tiny $0.8$}
\psfrag{1a}{\hspace{-0.15cm}\tiny $1$}
\psfrag{10e+0}{\hspace{-0.1cm}\tiny $10^0$}
\psfrag{10e+1}{\hspace{-0.1cm}\tiny $10^1$}
\psfrag{10e+2}{\hspace{-0.1cm}\tiny $10^2$}
\psfrag{0e0}{\hspace{-0.1cm}\tiny $ $}
\psfrag{1e3}{\hspace{-0.3cm}\tiny $10^5$}
\psfrag{1e4}{\hspace{-0.3cm}\tiny $10^6$}
\includegraphics[width=.8\textwidth]{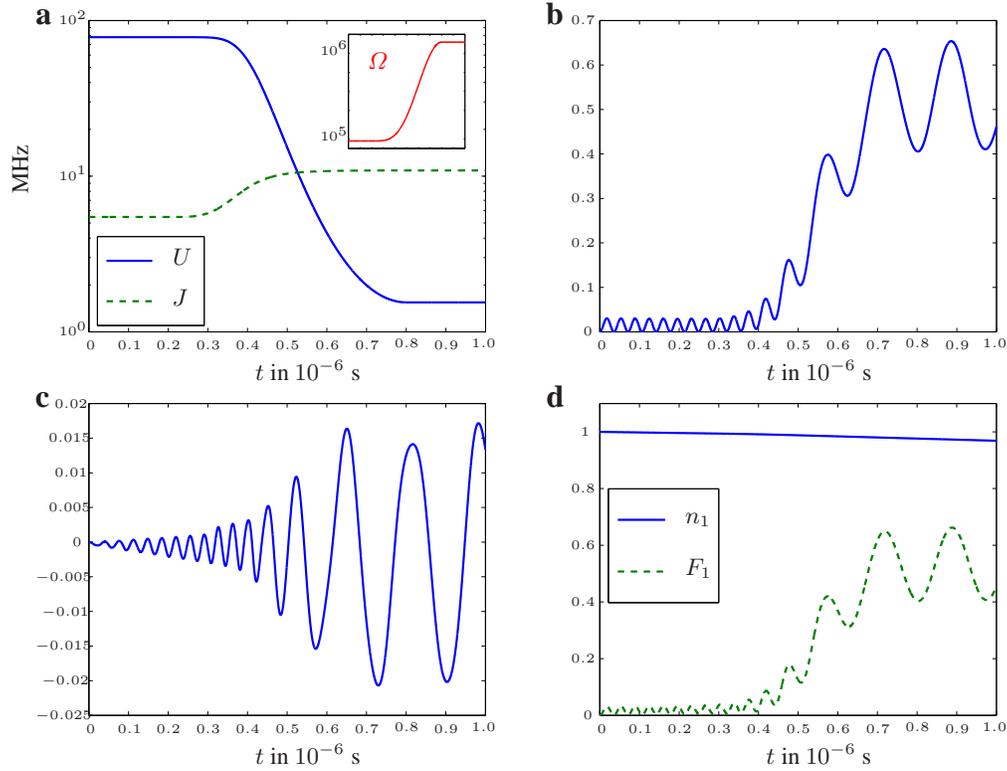}
\caption{\label{vardiff4} The Mott insulator to superfluid transition for 3 polaritons in 3 cavities compared to 3 particles in a 3 site Bose-Hubbard model.
{\bf a:} Log-plot of $U$ and $J$ and linear plot of the time dependent $\Omega$ (inset).
{\bf b:} Number fluctuations for polaritons in cavity 1, $F_1$, for a single quantum jump trajectory.
{\bf c:} Difference between number fluctuations for polaritons in a cavity and number fluctuations in a pure Bose-Hubbard model, $\delta F_1$, for a single quantum jump trajectory.
{\bf d:} Expectation value and fluctuations for the number of polaritons in one cavity/site according to
the effective model (\ref{bosehubbard}) with damping (\ref{gamma0}).
$g_{24} = g_{13} = 2.5 \times 10^9 s{-1}$, $\gamma_4 = \gamma_3 = 1.6 \times 10^7 s^{-1}$, $\kappa = 0.4 \times 10^5 s^{-1}$, $N = 1000$,
$\Delta = -2.0 \times 10^{10} \text{s}^{-1}$ and $2 \omega_C \alpha = 1.1 \times 10^{7} \text{s}^{-1}$.
The Rabi frequency of the driving laser is increased from initially $\Omega = 7.9 \times 10^{10} \text{s}^{-1}$ to finally $\Omega = 1.1 \times 10^{12} \text{s}^{-1}$.
Deviations from the pure Bose-Hubbard model are about 2\%. Reprinted with permission from \cite{HBP06}.}
\end{figure*}

Figure \ref{vardiff4} shows numerical simulations of the full dynamics of
an array of three cavities as described by equation (\ref{polariton_raw}), including spontaneous emission and cavity decay, and studies the number of polaritons in one cavity/site, 
$n_l = \langle p_l^{\dagger} p_l \rangle$ for site $l$, and the number fluctuations,
$F_l = \langle (p_l^{\dagger} p_l )^2 \rangle - \langle p_l^{\dagger} p_l \rangle^2$.
A comparison between the full model (\ref{polariton_raw}) and the effective model (\ref{bosehubbard}) is done by considering the differences
in the occupation numbers, $\delta n_l = \left[n_l\right]_{\text{cavities}} - \left[n_l\right]_{\text{BH}}$,
and number fluctuations, $\delta F_l = \left[F_l\right]_{\text{cavities}} - \left[F_l\right]_{\text{BH}}$, at each time step \cite{HBP06}.
Initially there is exactly one polariton in each cavity and parameters for toroidal micro-cavities from \cite{Spil05}, $g_{24} = g_{13} = 2.5 \times 10^9 s{-1}$, $\gamma_4 = \gamma_3 = 1.6 \times 10^7 s^{-1}$ and $\kappa = 0.4 \times 10^5 s^{-1}$, are assumed.
As the system is driven from a Mott insulator state to a superfluid state by ramping up the driving laser $\Omega$,
the particle number fluctuations increase significantly. 
The numerics in figure \ref{vardiff4} show good agreement of the full dynamics as described by equation (\ref{polariton_raw}) and the dynamics of the corresponding Bose-Hubbard model (\ref{bosehubbard}) and thus confirm the possibility of observing the Mott insulator-to-superfluid transition in such a system.
The oscillatory behaviour is related to the fact that the initial state of the system is, due to the
nonzero tunnelling rate $J$, not its ground state. This initial state has been chosen since it's preparation in an experiment is expected to be easier compared to other states and figure \ref{vardiff4} thus shows the dynamics that is expected to be observed in an experiment.

The phase diagram for this model and its dependence on the atom number $N$ have been analysed in \cite{RF07}, where also a glassy phase has been predicted for a system with disorder, see section \ref{sec:ph_diag} for details.

A polariton model which generates an effective Lieb-Liniger Hamiltonian and is
capable of reproducing the Tonks Girardeau regime was proposed in \cite{CGM+07}.
The scheme employs atoms of the same level structure as in figure \ref{level} that
couple to light modes of a tapered optical fibre. Furthermore quantum phase transitions of Dicke-model in cavity QED have been analysed \cite{DEPC07}.

\subsubsection{Two component Bose-Hubbard model}

Before we conclude this section, let us briefly review the proposal of Ref. \cite{HBP07c} for creating a two component polaritonic Bose-Hubbard model. This model, which describes the dynamics of two independent bosonic particles in each site, reads 
\begin{eqnarray} \label{bosehubbardtwocomp}
H_{\text{eff}} & = &
\sum_{k,j = b,c} \mu_j \, n^{(j)}_{k} \, - \, \sum_{< k, k' >; j,l = b,c} 
J_{j,l} \left( j_{k}^{\dagger} \, l_{k'}\, + \, \text{h.c.} \right) \, \nonumber \\
& + & \sum_{k,j = b,c} U_j \, n^{(j)}_{k} \left(n^{(j)}_{k} - 1\right) \, 
+ \sum_{k} U_{b,c} \, n^{(b)}_{k} n^{(c)}_{k} \, ,
\end{eqnarray}
where $b_{k}^{\dagger}$($c_{k}^{\dagger}$) create polaritons of the type $b$($c$) in the cavity at site $k$,  $n^{(b)}_{k} = b_{k}^{\dagger} b_{k}$ and
$n^{(c)}_{k} = c_{k}^{\dagger} c_{k}$. $\mu_b$ and $\mu_c$ are the polariton energies, $U_b$, $U_c$ and $U_{b,c}$ their on-site interactions and $J_{b,b}$, $J_{c,c}$ and $J_{b,c}$ their tunnelling rates. This model exhibits interesting quantum many-body phenomena such as spin density separation \cite{RFZZ03}, spin order in the Mott regime \cite{DDL03} and phase separation \cite{CH03}.

The Hamiltonian (\ref{bosehubbardtwocomp}) can be realised in the same set-up as above, by considering the atom-photon interactions in a dispersive regime with $\delta \gg \Omega, g$.
Then, the polariton operators (\ref{polariton_operators}) read, 
\begin{equation} \label{polariton_operators_2}
\begin{array}{rlrl}
p_0^{\dagger} = & \frac{1}{B} \, \left(g S_{12}^{\dagger} - \Omega a^{\dagger} \right) & \quad \mu_0 = & 0 \nonumber \\
p_-^{\dagger} \approx & \frac{1}{B} \, \left(\Omega S_{12}^{\dagger} + g a^{\dagger} \right) - \frac{B}{\delta} S_{13}^{\dagger} & \quad \mu_- = & - \frac{B^2}{\delta} \nonumber \\
p_+^{\dagger} \approx & S_{13}^{\dagger} + \frac{1}{\delta} \, \left(\Omega S_{12}^{\dagger} + g a^{\dagger} \right) & \quad \mu_+ = & \delta + \frac{B^2}{\delta}
\end{array}\,, 
\end{equation}
to leading order in $\delta^{-1}$. Note that now, not only the polariton $p_0$, but also $p_-^{\dagger}$ do not experience loss from spontaneous emission to leading order in $\delta^{-1}$. One can therefore define two dark state polaritons species
\begin{equation} \label{b_c_def}
b^{\dagger}  = \frac{1}{B} \, \left(g S_{12}^{\dagger} - \Omega a^{\dagger} \right) \, ; \quad
c^{\dagger} = \frac{1}{B} \, \left(\Omega S_{12}^{\dagger} + g a^{\dagger} \right) \, .
\end{equation}
From an analysis similar to the one outlined above for the one component case, it can be shown that the dynamics of these two polaritons species is given by the two component Bose-Hubbard model (\ref{bosehubbardtwocomp}), where the parameters are given by,
$J_{bb} = \alpha \frac{g^2}{B^2}$,
$J_{cc} = \alpha \frac{\Omega^2}{B^2}$,
$J_{bc} = \alpha \frac{g \Omega}{B^2}$,
$U_b = - \frac{g_{24}^2 g^2 \Omega^2}{B^4 \Delta}$,
$U_c = - \frac{g_{24}^2 g^2 \Omega^2}{B^4 (\Delta + 2 B^2/\delta)}$ and
$U_{bc} = - \frac{g_{24}^2 (g^2 - \Omega^2)^2}{B^4 (\Delta + B^2/\delta)}$.
For further details of the derivation the reader is referred to Ref. \cite{HBP07c}.

\begin{figure}
\centering
\psfrag{xa}{\hspace{0.1cm} $\frac{\Omega}{g_{13}}$}
\psfrag{xb}{\hspace{0.1cm} $\frac{\Omega}{g_{13}}$}
\psfrag{a1}{\raisebox{-0.2cm}{\scriptsize $10^{1}$}}
\psfrag{a2}{\raisebox{-0.2cm}{\scriptsize $10^{2}$}}
\psfrag{a3}{\raisebox{-0.2cm}{\scriptsize $10^{3}$}}
\psfrag{aa1}{\hspace{-0.5cm} \raisebox{-0.0cm}{\scriptsize $10^{-3}$}}
\psfrag{aa2}{\hspace{-0.5cm} \raisebox{-0.0cm}{\scriptsize $10^{-2}$}}
\psfrag{aa3}{\hspace{-0.5cm} \raisebox{-0.0cm}{\scriptsize $10^{-1}$}}
\psfrag{aa4}{\hspace{-0.3cm} \raisebox{-0.0cm}{\scriptsize $10^{0}$}}
\psfrag{aa5}{\hspace{-0.3cm} \raisebox{-0.0cm}{\scriptsize $10^{1}$}}
\psfrag{b1}{\raisebox{-0.2cm}{\scriptsize $10^{1}$}}
\psfrag{b2}{\raisebox{-0.2cm}{\scriptsize $10^{2}$}}
\psfrag{b3}{\raisebox{-0.2cm}{\scriptsize $10^{3}$}}
\psfrag{bb1}{\hspace{-0.5cm} \raisebox{-0.0cm}{\scriptsize $10^{-4}$}}
\psfrag{bb2}{\hspace{-0.5cm} \raisebox{-0.0cm}{\scriptsize $10^{-3}$}}
\psfrag{bb3}{\hspace{-0.5cm} \raisebox{-0.0cm}{\scriptsize $10^{-2}$}}
\psfrag{bb4}{\hspace{-0.5cm} \raisebox{-0.0cm}{\scriptsize $10^{-1}$}}
\psfrag{bb5}{\hspace{-0.3cm} \raisebox{-0.0cm}{\scriptsize $10^{0}$}}
\psfrag{bb6}{\hspace{-0.3cm} \raisebox{-0.0cm}{\scriptsize $10^{1}$}}
\includegraphics[width=.4\linewidth]{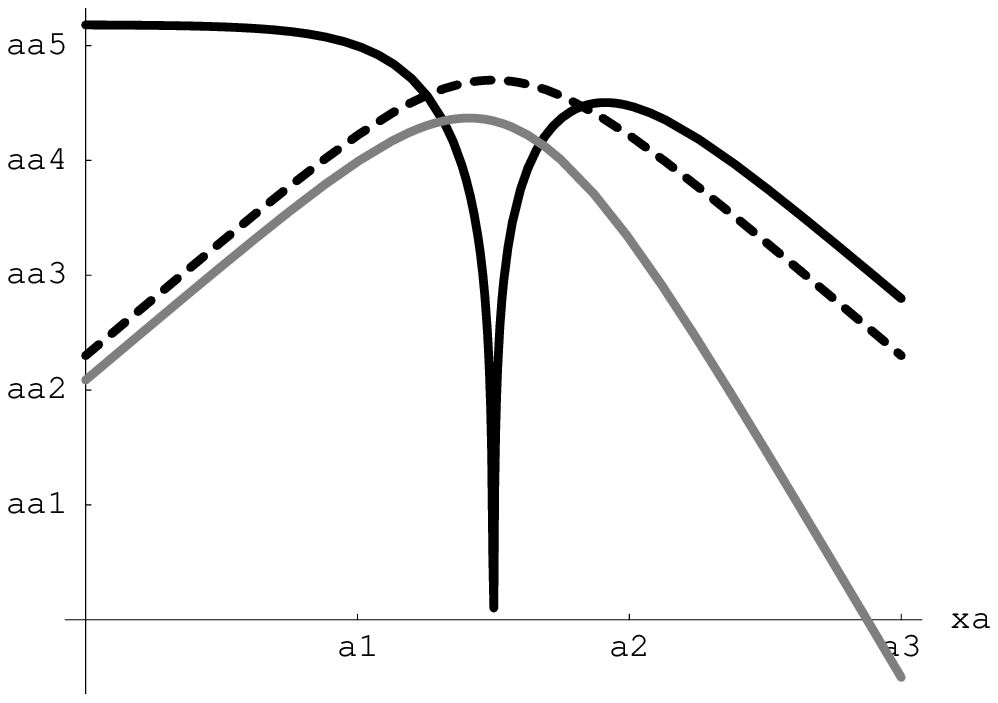}
\hspace{.1\linewidth}
\includegraphics[width=.4\linewidth]{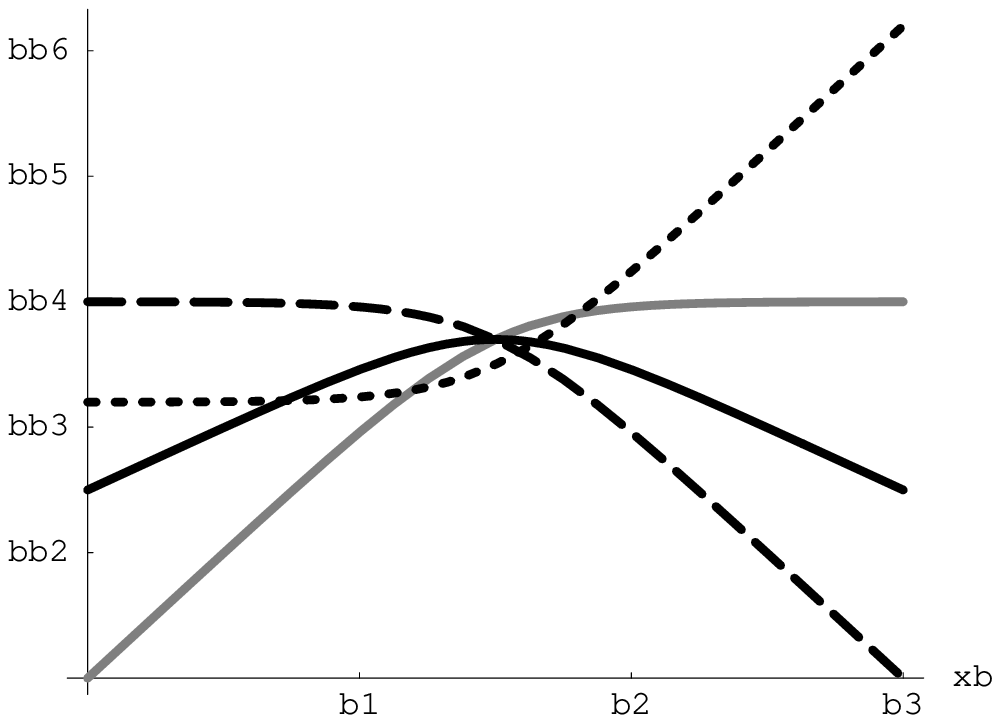}
\caption{Left: The polariton interactions $U_b$ (dashed line), $U_c$ (grey line) and $U_{bc}$
(solid line) in units of $g_{13}$ as a function of $\Omega / g_{13}$. Right: The tunnelling rates $|J_{bb}|$ (dashed line), $|J_{cc}|$ (grey line) and $|J_{bc}|$ (solid line) together with $|\mu_c - \mu_b|$ (dotted line) in units of $g_{13}$ as a function of $\Omega / g_{13}$. The parameters of the system are $g_{24} = g_{13}$, $N = 1000$, $\Delta = - g_{13} / 20$, $\delta = 2000 \sqrt{N} g_{13}$ and $\alpha = g_{13} / 10$. Reprinted with permission from \cite{HBP07c}.}
\label{range_HBP07c}
\end{figure}

An example how the parameters of the effective Hamiltonian (\ref{bosehubbardtwocomp})
vary as a function of the intensity of the driving laser $\Omega$ is given in figure
\ref{range_HBP07c}, where the parameters of the atom cavity system are chosen to be
$g_{24} = g_{13}$, $N = 1000$, $\Delta = - g_{13} / 20$, $\delta = 2000 \sqrt{N} g_{13}$ and $\alpha = g_{13} / 10$. Figure \ref{range_HBP07c} shows the interactions $U_b$, $U_c$ and $U_{bc}$, the tunnelling rates $J_{bb}$, $J_{cc}$ and $J_{bc}$ and $|\mu_c - \mu_b|$ as a function of $\Omega / g_{13}$.
For $g \approx \Omega$ one has $|U_{bc}| \ll |U_b|, |U_c|$ and $J_{bb} \approx J_{cc} \approx J_{bc}$.
Whenever $|\mu_c - \mu_b| < |J_{bc}|$, $b^{\dagger}$ polaritons get converted into $c^{\dagger}$ polaritons
and vice versa via the tunnelling $J_{bc}$.
With the present choice of $\alpha$ and $\delta$, this happens for
$0.16 g < \Omega <  1.6 g$.

\subsection{Frozen light: photonic Bose-Hubbard models}

In the last section we have discussed that, as soon as a strong coupling regime with a cooperativity factor larger than unity, $\xi \gg 1$ is achieved, it is possible to engineer polaritonic Bose-Hubbard models in coupled cavity arrays. The prospect that even higher cooperativity factors (of the order of $10^2-10^3$ \cite{Spil05}) can be realised in some cavity QED setups, opens up the possibility of engineering a Bose-Hubbard model also for the cavity mode photons. Such a setting is actually conceptually simpler than the polaritonic and has already been outlined before. Given that the photonic tunnelling term is a consequence of the interaction of neighbouring cavities, to realise a photonic Bose-Hubbard model one needs to find a way to efficiently create large photon Kerr nonlinearities using the interaction of the atoms with the cavity mode. In this section we review two proposals to this aim.

To create a pure photonic Bose-Hubbard model would be very interesting as in this way new regimes for light, which do not occur naturally, could be engineered. For example, in the photonic Mott insulator exactly one photon exists in each cavity,
provided the whole structure contains on average one photon per cavity. Moreover, the photons are localised in the cavity they are in and are not able to hop between different cavities. In such a situation photons behave as strongly correlated particles that are each "frozen" to their lattice site, a system that would correspond to a crystal formed by light.

The first proposal for generating strong Kerr photon nonlinearities, which we discuss here, is the EIT based setting analysed in the last section. As shown in \cite{HP07}, in the regime characterised by $g \ll \Omega$ and $g_{24}g \ll |\Delta \Omega|$, the dark sate polariton $p_0$ becomes a photon and the nonlinearity, c.f. equation (\ref{osint}), is given by 
\begin{equation}
U = -  \frac{g_{24}g}{|\Delta \Omega|}  \frac{g}{\Omega} g_{24} (a^{\cal y})^2 a^2. 
\end{equation}
The effective decay rate for the photons subject to the nonlinearity is given by
\begin{equation}
\Gamma = \kappa + \Theta(n - 2)g_{24}^2g^2\Delta^{-2}\Omega^{-2}\gamma_4 \, ,
\end{equation}
where cavity decay is the main source of loss since $g_{24}g \ll |\Delta \Omega|$.

\begin{figure}
\psfrag{data1}{\hspace{0.0cm} \small $U$}
\psfrag{data2}{\hspace{0.0cm} \small $J$}
\psfrag{a}{\hspace{-2.0cm}\raisebox{-0.0cm}{{\bf a}}}
\psfrag{b}{\hspace{-2.0cm}\raisebox{-0.0cm}{{\bf b}}}
\psfrag{c}{\hspace{-2.0cm}\raisebox{-0.0cm}{{\bf c}}}
\psfrag{d}{\hspace{-2.0cm}\raisebox{-0.0cm}{{\bf d}}}
\psfrag{t}{\hspace{-0.4cm}\raisebox{-0.1cm}{\tiny $t$ in $10^{-6}$ s}}
\psfrag{n}{\tiny $n_1$}
\includegraphics[width=\linewidth]{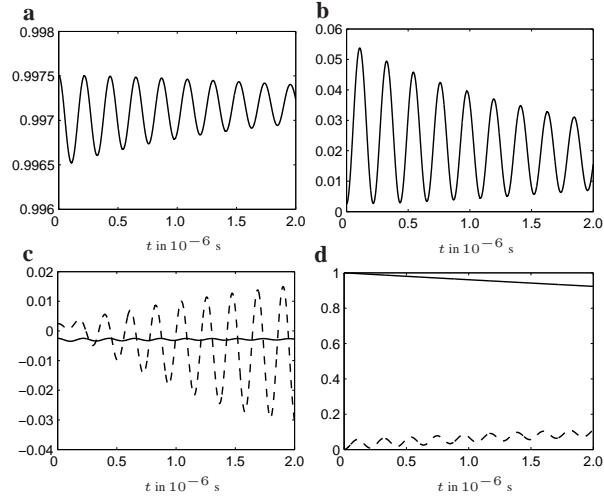}
\caption{\label{photonMott} Mott insulator state for three photons in three cavities:
{\bf a}: photon number $n_1$, {\bf b}: photon number fluctuation $F_1$, {\bf c}: differences between full and effective Bose-Hubbard model, $\delta n_1$ (solid line) and $\delta F_1$ (dashed line) and {\bf d}: $n_1$ and $F_1$ as given by a solution of a master equation for the three site Bose-Hubbard model (\ref{bosehubbard}) including damping $\Gamma$. Reprinted with permission from \cite{HP07}.}
\end{figure}

Taking, for example, the case of toroidal micro-cavities, for which the achievable parameters are predicted to be $g_{24} = 2.5 \times 10^9 s^{-1}$, $\gamma_4 = 1.6 \times 10^7 s^{-1}$, and $\kappa = 0.4 \times 10^{5} s^{-1}$ \cite{Spil05}, and setting the parameters such that $g/\Omega = g_{24}g / |\Delta \Omega| = 0.1$, leads to $U/\Gamma = 625$, which would be sufficient to observe a Mott insulator of photons, as shown in figure \ref{photonMott}.

\begin{figure}
  \sidecaption
  \includegraphics[width=.5\linewidth]{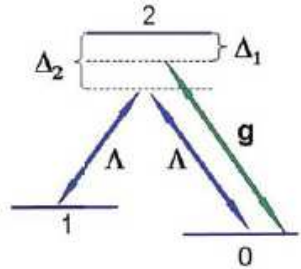}
  \caption{The level structure for the generation of the effective Hamiltonian (\ref{h2}). ($g$, $\Delta_1$) and $(\Lambda$, $\Delta_2$) are the Rabi frequencies and detunings of the cavity-atom and laser-atom interactions. Reprinted with permission from \cite{HBP07b}.}
  \label{levels_fernando}
\end{figure}

A second proposal to realise large Kerr nonlinearities is due to Ref. \cite{HBP07b}. There a new configuration, involving a three levels lambda structure and only one coupling to the cavity mode (see figure \ref{levels_fernando}), was shown to produce nonlinearities as large as the EIT based setting discussed before. A possible advantage of this approach is that, as the only three levels are used, it might be easier to implement it in some of the new cavity QED systems, such as quantum dots in photonic crystal cavities or Cooper pair boxes in stripline cavities.

The relevant atomic level structure, depicted in figure \ref{levels_fernando}, is a $\Lambda$ system with two metastable states and an excited state. The cavity mode couples dispersively only to the $0-2$ transition and the levels $0$ and $1$ are coupled via a far-detuned Raman transition. Finally, the detuning associated with the lasers and with the cavity mode are assumed to be sufficiently different from each other. The origin of the nonlinearity can be understood entirely by considering the Stark shifts due to the dispersive interactions.

The full Hamiltonian of the system, in the interaction picture, reads
\begin{equation} \label{hfull}
H = g e^{- i \Delta_1 t} a S_{20} + \sqrt{2} \Lambda e^{- i \Delta_2 t} S_{2+} + \text{h.c.}
\end{equation}
where $S_{20} := \sum_k \ket{2_k}\bra{0_k}$ and  $S_{2+} := \sum_k \ket{2_k}\bra{+_k}$. As shown in figure \ref{levels_fernando}, ($g$, $\Delta_1$) and $(\Lambda$, $\Delta_2$) are the Rabi frequencies and detunings of the cavity-atom and laser-atom interactions, respectively. In the dispersive regime, characterised by $\frac{\sqrt{N} g}{\Delta_1} \ll 1$ and $\frac{ \sqrt{N} \Lambda}{\Delta_2} \ll 1$, where furthermore $\sqrt{N}g, \sqrt{N} \Lambda \ll |\Delta_2 - \Delta_1|$, the processes driven by the cavity-atom and laser-atom interactions can be treated independently and the excited state will hardly be populated. Hence an effective Hamiltonian for the two metastable states and the cavity mode can be found by adiabatically eliminating the excited state. Dropping out terms proportional to the identity, it is given by
\begin{equation}
H_1 = \frac{g^2}{\Delta_1}a^{\cal y}a S_{00} + \frac{\Theta}{2} (S_{10} + S_{01}),
\end{equation}
with $\Theta := 2\Lambda^2/\Delta_2$. 

In a second interaction picture with respect to $H_0 = \frac{g^2}{2\Delta_1}a^{\cal y}a$ it reads $H_1^{int} = \frac{g^2}{2\Delta_1} a^{\cal y}a \left( S_{+-} + S_{-+}\right) + \frac{\Theta}{2} S_3$, where $S_{+-} := \sum_k \ket{+_k}\bra{-_k}$, $S_{-+} = (S_{+-})^{\cal y}$, and, as before, $S_3 := \sum_{k=1}^N \ket{+_k}\bra{+_k} - \ket{-_k}\bra{-_k}$. In the $\{ \ket{\pm} \}$ basis the system can thus be viewed as an ensemble of two level atoms driven by a laser with a photon-number-dependent Rabi frequency. Considering the dispersive regime of this system, $\frac{\sqrt{N}g^2}{2\Delta_1 \Theta} \ll 1$, the atoms prepared in the $\ket{-}$ state will experience a Stark shift proportional to $(a^{\cal y}a)^2$, which gives rise to a strong Kerr nonlinearity. The effective Hamiltonian is be given by $H_{eff} = \frac{g^4}{4\Delta_1^2\Theta} a^{\cal y}a a^{\cal y}a S_3$. Therefore, if all atoms are prepared in the $\ket{-}$ state, one obtains an essentially absorption free Kerr nonlinearity given by
\begin{equation} \label{h2}
H_{kerr} = \frac{\sqrt{N}g^2}{2\Delta_1 \Theta} \frac{\sqrt{N} g}{2 \Delta_1} g a^{\cal y}a a^{\cal y}a.
\end{equation}

\subsection{Measurements \label{sec:measure}}

In this section we discuss ways how information about the state of the polaritonic
Bose-Hubbard models can be obtained in experiments. The possibilities to access the system in measurements are somewhat complementary to those for cold atoms in optical lattices \cite{BDZ08}. Whereas time of flight images give access to global quantities
such as phase coherence in optical lattices, coupled cavity arrays allow to access
the local particle statistics in measurements. Unlike optical lattices they would thus allow to obtain direct experimental evidence for the exact integer particle number
in each lattice site in a Mott Insulator regime. We first discuss the measurement scheme for the one-component model and then turn to its extension to for the two-component model.

\subsubsection{One-component model}

For the model (\ref{bosehubbard}), the number of polaritons in one individual cavity can be measured via state selective resonance fluorescence. To that end, the polaritons are transformed into purely atomic excitations by adiabatically switching off the driving laser. The process is adiabatic if $g B^{-2} \frac{d}{dt} \Omega \ll |\mu_+| , |\mu_-|$ \cite{FIM05}, which means it can be fast enough to prevent polaritons from hopping between cavities during the switching. Hence, in each cavity, the final number of atomic excitations in level 2 is equal to the initial number of dark state polaritons, while the other polariton species will not be transferred to the atomic level 2 (c.f. (\ref{polariton_operators})). In this state cavity loss is eliminated and the atomic excitations are very long lived. They can then be measured with high efficiency and precision using state-selective resonance fluorescence as proposed in \cite{I02} and \cite{JK02}.

\subsubsection{Two-component model}

The number statistics for both polariton species for the model (\ref{bosehubbardtwocomp}), $b^{\dagger}$ and $c^{\dagger}$, in one cavity can again be
measured using state selective resonance fluorescence. In the two-component case the STIRAP can however not be applied as in the single component case because the energies $\mu_b$ and $\mu_c$ are similar and the passage would thus need to be extremely slow to be adiabatic.

\begin{figure}
	\centering
	\psfrag{A}{\bf a}
	\psfrag{B}{\bf b}
	\psfrag{g24}{$g_{24}$}
	\psfrag{o}{\hspace{-0.1cm}$\Theta$}
	\psfrag{L1}{$\Lambda$}
	\psfrag{L2}{$\Lambda$}
	\psfrag{d}{\hspace{-0.1cm}$\Delta$}
	\psfrag{d2}{\hspace{-0.0cm}$\delta_{\Lambda}$}
	\psfrag{e}{$\varepsilon$}
	\psfrag{w}{\hspace{-0.2cm}$\omega_C$}
	\psfrag{1}{\raisebox{-0.1cm}{$1$}}
	\psfrag{2}{\raisebox{-0.14cm}{$2$}}
	\psfrag{3}{$3$}
	\psfrag{4}{$4$}
	\raisebox{.06\linewidth}{\includegraphics[width=.36\linewidth]{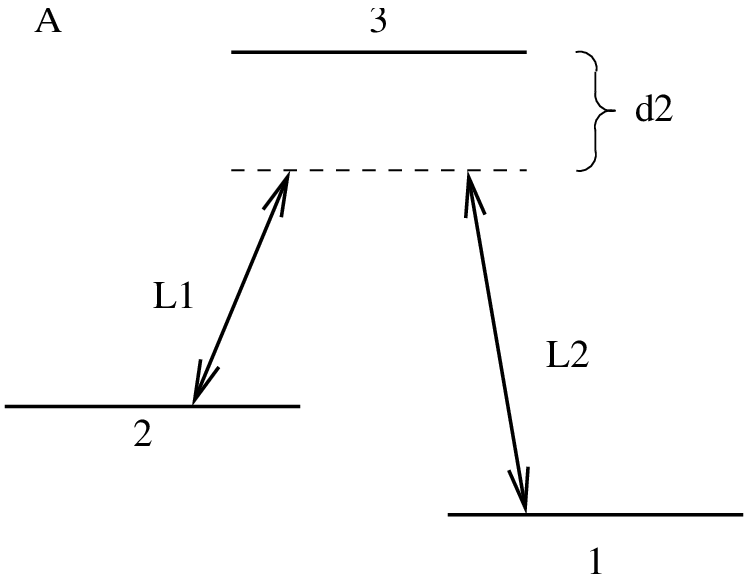}}
        \hspace{.06\linewidth}
	\includegraphics[width=.48\linewidth]{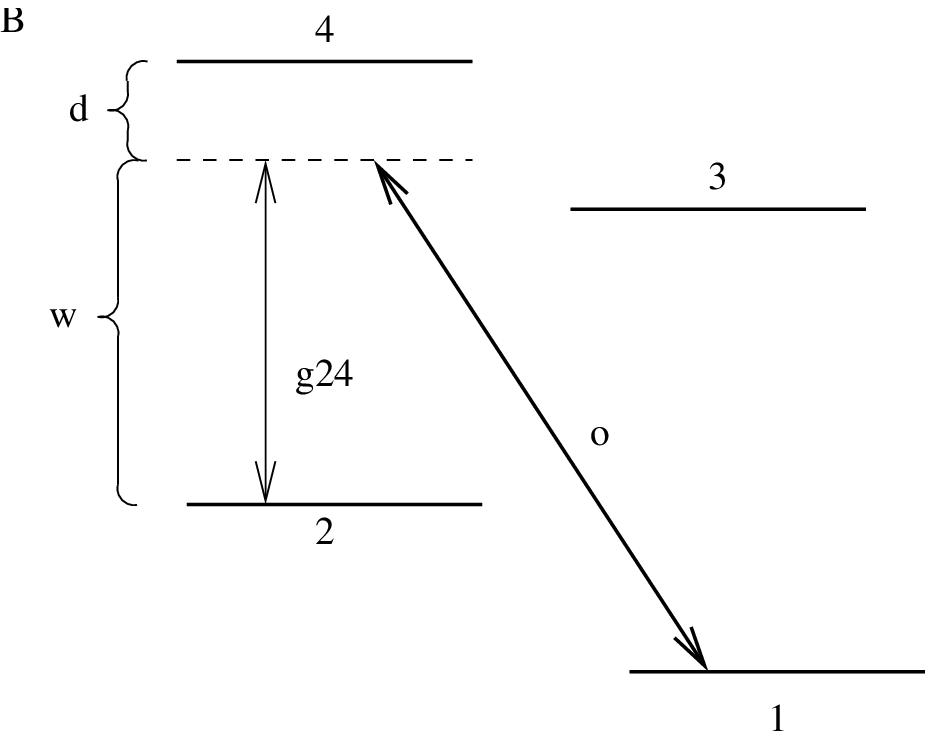}
	\caption{{\bf a}: Configuration of the $\pi / 2$-pulse. Two driving lasers in two-photon transition with identical Rabi frequencies $\Lambda$ couple to the atomic transitions $1 \leftrightarrow 3$ and $2 \leftrightarrow 3$. {\bf b}: Configuration for the STIRAP process. A driving laser couples to the $1 \leftrightarrow 4$ transition with Rabi frequency $\Theta$. The cavity mode couples to transitions $2 \leftrightarrow 4$ and $1 \leftrightarrow 3$, where the coupling to $1 \leftrightarrow 3$ is ineffective and not shown. Reprinted with permission from \cite{HBP07c}.}
	\label{flippass}
\end{figure}

For two components, one can do the measurements as follows. First the external driving laser $\Omega$ is switched off. Then the roles of atomic levels 1 and 2 are interchanged in each atom via a Raman transition by applying a $\pi / 2$-pulse. To this end the transitions $1 \leftrightarrow 3$ and $2 \leftrightarrow 3$ are driven with two lasers (both have the same Rabi frequency $\Lambda$) in two-photon resonance for a time
$T = \pi \delta_{\Lambda} / |\Lambda|^2$ ($\delta_{\Lambda}$ is the detuning from atomic level 3). The configuration is shown in figure \ref{flippass}a. This pulse results in the mapping
$\ket{1_j} \leftrightarrow \ket{2_j}$ for all atoms $j$.

Next another laser, $\Theta$, that drives the transition
$1 \leftrightarrow 4$ is switched on, see figure \ref{flippass}b. Together with the coupling $g_{24}$, this configuration can be described in terms of three polaritons, $q_0^{\dagger}$, $q_+^{\dagger}$ and $q_-^{\dagger}$, in an analogous way to
$p_0^{\dagger}$, $p_+^{\dagger}$ and $p_-^{\dagger}$, where now the roles of the atomic
levels 1 and 2 and the levels 3 and 4 are interchanged.
Hence, if one chooses $\Theta = \Omega$ the $\pi / 2$-pulse maps the $b^{\dagger}$ onto the dark state polaritons of the new configuration, $q_0^{\dagger}$, whereas
for $\Theta = - \Omega$ it maps the $c^{\dagger}$ onto $q_0^{\dagger}$. The driving laser is then adiabatically switched off, $\Theta \rightarrow 0$, and the corresponding STIRAP process maps the $q_0^{\dagger}$ completely onto atomic excitations of level 1. This process can be fast since the detuning $\Delta$ is significantly smaller than $\delta$ and hence the energies of all polariton species $q_0^{\dagger}$, $q_+^{\dagger}$ and $q_-^{\dagger}$ are well separated. Another $\pi / 2$-pulse finally maps the excitations of level 1 onto excitations of level 2,
which can be measured by state selective resonance fluorescence in the same way as for the one-component model.

The whole sequence of $\pi / 2$-pulse, STIRAP process and another $\pi / 2$-pulse can be done much faster than the timescale set by the dynamics of the Hamiltonian (\ref{bosehubbardtwocomp}) and $b^{\dagger}$ or $c^{\dagger}$ can be mapped onto atomic excitations in a time in which they are not able to move between sites. 
The procedure thus allows to measure the instantaneous local particle statistics of each species separately.
%


\section{Models Based on the Jaynes-Cummings Hamiltonian} \label{sec:JC}

Many-body dynamics on a lattice that can either be dominated by the tunnelling between adjacent
lattice sites or by an on-site interaction can also be generated in an array of coupled cavities with
Jaynes-Cummings type coupling between cavity photons and two level atoms in each cavity.
As shown in \cite{ASB07} and \cite{GTCH06}, the anharmonicity of the energy levels of the dressed states of a Jaynes-Cummings Hamiltonian acts as an effective on-site interaction. To see this let
us recall the Jaynes-Cummings Hamiltonian for one atom,
\begin{equation} \label{JC_JCH}
H^{JC} = \omega_C a^{\cal y}a + \omega_0 \ket{e}\bra{e} + g (a^{\cal y}\ket{g}\bra{e} + a\ket{e}\bra{g}),
\end{equation}
where $\omega_C, \omega_0$ are the frequencies of the resonant mode of the cavity and of the atomic transition, respectively, $g$ is Jaynes-Cummings coupling between the cavity mode and the two level system, $a^{\cal y}$ is the creation operator of a photon in the resonant cavity mode, and $\ket{g}, \ket{e}$ are the ground and excited states of the two level system.

Since the energy of a photon $\omega_C$ and the atomic transition energy $\omega_0$
are much greater than the atom photon coupling $g$, the number of excitations is conserved
for the Hamiltonian (\ref{JC_JCH}). Hence it can be diagonalised for each manifold with a fixed number of excitations $n$ separately. The energy eigenvalues for $n$ excitations read $E_0 = 0$
and
\begin{equation} \label{JC_dressed}
E_n^{\pm} =  n \omega_C + \frac{\Delta}{2} \pm \sqrt{n g^2 + \frac{\Delta^2}{4}}
\end{equation}
for $n \ge 1$, where $\Delta = \omega_0 - \omega_C$.

\begin{figure}
 \centering
 \psfrag{dE}{$(E_2^- - 2 E_1^-) / g$}
 \psfrag{x}{$\Delta / g$}
 \includegraphics[width=.6\linewidth]{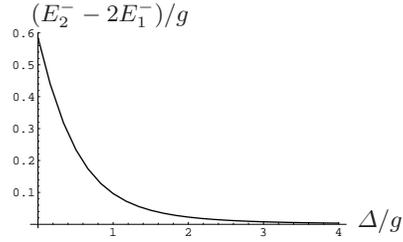}
 \caption{\label{JC_rep_fig} Dependence of the effective repulsion $E_2^- - 2 E_1^-$ on the detuning $\Delta$. The effective repulsion vanishes for large $\Delta$}
\end{figure}

Let us now consider an array of cavities where the atom photon interaction in each cavity is described by equation (\ref{JC_JCH}) and assume that we have prepared the system in a state with
one excitation of energy $E_1^-$ in each cavity. Since the lowest energy for two excitations in one cavity is $E_2^-$, moving one additional excitation to a cavity requires an extra energy of
\begin{equation} \label{effJCrep}
E_2^- - 2 E_1^- = 2 \sqrt{g^2 + \frac{\Delta^2}{4}} - \sqrt{2 g^2 + \frac{\Delta^2}{4}} - \frac{\Delta}{2} ,
\end{equation}
which plays the role of an effective on-site repulsion. This repulsion can be tuned via the detuning $\Delta$ as figure \ref{JC_rep_fig} shows.

The full effective many-body model in a cavity array that is based on the effective interaction
(\ref{effJCrep}) reads,
\begin{equation} \label{JC_eff_mb}
 H = \sum_{\vec{R}} H^{JC}_{\vec{R}}
- J \sum_{<\vec{R}, \vec{R}'>} \left( a_{\vec{R}}^{\dagger} a_{\vec{R}'} + \text{h.c.} \right),
\end{equation}
where $\vec{R}$ labels the site of a cavity. Each cavity contains one atom interacting via the Jaynes-Cummings interaction with the cavity mode and photons tunnel between neighbouring cavities at a rate $J$.

\begin{figure}
        \centering
	\includegraphics[width=.8\linewidth]{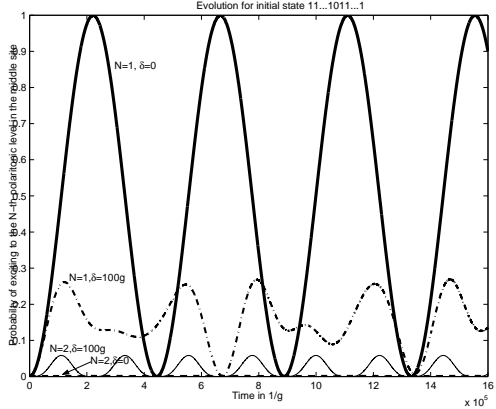}
	\caption{Probabilities for finding one (thick solid line) or two
        (thick dashed line) excitations in the middle cavity which was
        initially taken to be empty (in the resonant regime).
        In this case, as there were initially fewer excitations than
        cavities, oscillations occur for the single excitation.
        The double one is never occupied due to
	the blockade effect-Mott insulator phase (thick dashed line).
        For the detuned case however, hopping of more than one excitation is
	allowed (superfluid regime) which is evident as the higher
	excitation manifolds are being populated now (dashed line and thin
	solid line). Reprinted with permission from \cite{ASB07}.} \label{pop_ASB06}
\end{figure}

Figure \ref{pop_ASB06} gives evidence for the Mott insulator and superfluid regimes
of Hamiltonian (\ref{JC_eff_mb}). The probabilities to find one or two excitations in one cavity are shown for an array where initially one cavity had no and all others one excitation. For detuning $\Delta = 0$, double occupations never occur, indicating a Mott insulator behaviour. For detunings $\Delta \gg g$, on the other hand, occupation numbers larger than one occur as to be expected for a superfluid regime. In the deep Mott insulator regime, the model (\ref{JC_eff_mb})
can also reproduce effective spin models \cite{ASB06}, see section \ref{sec:JCspin}.

Effective many-body physics based on the Hamiltonian (\ref{JC_eff_mb}) can not only be observed for a single two level system in each cavity, as assumed in (\ref{JC_JCH}), but also for setups with several two level systems per cavity. Such a model can describe photonic crystal micro-cavities doped with substitutional donor or acceptor impurities. This approach to implementing effective many-body models, which can have suitable parameters, has been proposed in \cite{Na08}. The phase transitions of a model with several two level systems in each cavity have also been studied in \cite{RF07,LL08}.

To allow for the observation of coherent effective many-body dynamics, the effective on-site repulsion (\ref{effJCrep}) has to be much stronger than the rates of spontaneous emission and cavity decay. Since the dressed states are superpositions of a photonic and an atomic component, they are vulnerable to both decay mechanisms. Spontaneous emission can be suppressed by increasing the detuning $\Delta$ as this causes the occupation of the excited atomic level to decrease.
An increase of $\Delta$ however also causes the repulsion to decrease as shown in figure \ref{JC_rep_fig}. The effective many-body dynamics is thus only observable if each cavity operates in the strong coupling regime, see section \ref{cQED}.

\begin{figure}
        \centering
	\includegraphics[width=\linewidth]{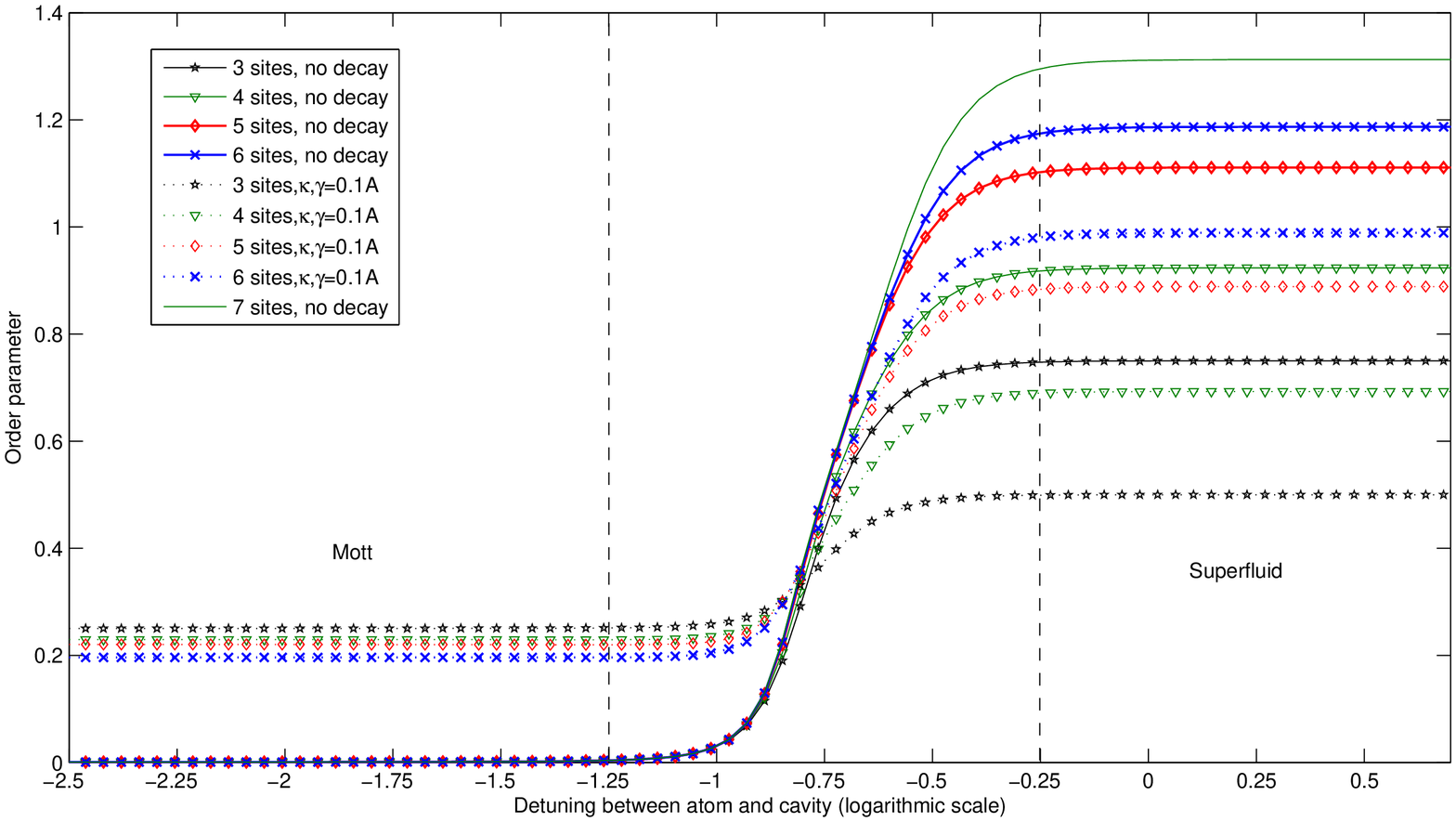}
	\caption{The order parameter $F_l$ as a function of the detuning between the hopping photon and the doped two-level system in logarithmic units of the matter-light coupling $g$. Simulations include results for 3–7 sites, with and without dissipation due to spontaneous emission and cavity leakage. The graphs are obtained by diagonalisation of the Hamiltonian (\ref{JC_eff_mb}).} \label{mott_ASB06}
\end{figure}

As for the polaritonic Bose-Hubbard model (section \ref{sec:pbh}), the transition between Mott insulator and superfluid phase can be characterised by the fluctuations of the number of excitations in one single cavity. Here, the number of excitations in cavity $l$ is measured by the operator ${\cal N}_l = a_l^{\dagger} a_l + \proj{e_l}$ and their fluctuations $F_l$ are given by the ground state variance of ${\cal N}_l$, $F_l = \langle {\cal N}_l^2 \rangle - \langle {\cal N}_l \rangle^2$. Figure \ref{mott_ASB06} shows the number fluctuations $F_l$, which play the role of an order parameter, as a function of the atom cavity detuning $\Delta$. Whereas the fluctuations are very small in the Mott insulator regime (small detuning), they become significantly larger in the superfluid phase (large detuning).

Various approaches have shown that this model indeed exhibits a quantum phase transition that closely resembles the Mott insulator to superfluid transition of the Bose-Hubbard model \cite{ASB07,GTCH06,RF07,Na08,AHTL08,MCT+08}, see also section \ref{sec:ph_diag}. On the other hand deviations of the phase transitions in the model (\ref{JC_eff_mb}) from the Mott insulator to superfluid transition in a Bose-Hubbard model have also been found.

In this analysis exact diagonalisations of small models of very few cavities have been performed \cite{IOK08}. Here, insulator states have been found, where the number of excitations in one site does not fluctuate and the effective particles are either atomic (for $\Delta / g < -1$ and $|J| < |\Delta|$) or polaritonic ($\Delta / g \lesssim 1$ and $|J| \lesssim |\Delta|$). Superfluid states on the other hand show local excitation number fluctuations and can either contain polaritonic
($\Delta / g < -1$ and $|J| \approx |\Delta|$) or photonic ($\Delta < 0, |J| > |\Delta| \gtrsim g$ or $\Delta > 0, |J| \gtrsim g$) effective particles.

For the characterisation of the various phases, possible signatures in the visibility of interference fringes of photons that could be emitted by the cavities have been discussed \cite{RF07,HLSS08}. The visibility ${\cal V}$ of the interference pattern is given in terms
of the photon number distribution in momentum space,
$ S(k) = \frac{1}{L} \sum_{j,l = 1}^L \ee^{2 \pi \iu (j - l)/L} \Bra a_j^{\dagger} a_l \Ket$,
($L$ is the number of sites in a one dimensional array) by
${\cal V} = \left( \text{max}(S) - \text{min}(S) \right) / \left( \text{max}(S) + \text{min}(S) \right)$. In the superfluid regime, ${\cal V}$ approaches unity whereas it is very small for a
Mott insulator.


\section{Phase Diagrams} \label{sec:ph_diag}

As already indicated above one of the key questions for effective many-body systems in coupled cavity arrays is which quantum phase transitions these can show.
The phase diagram of both, polaritonic Bose-Hubbard models and Jaynes-Cummings interaction based models, has been analysed in various settings. Approaches based on different techniques have
been carried out for one- and two-dimensional models.

The phase boundary between a Mott insulator and a superfluid phase can be determined in a
grand canonical approach. To this end, a chemical potential $\mu$ is introduced and the
grand canonical Hamiltonian reads,
\begin{equation} \label{H_GC}
 H^{GC} = H - \mu {\cal N} ,
\end{equation}
where the operator ${\cal N}$ counts the total number of excitations and is given by
\begin{equation}
 {\cal N} = \sum_{\vec{R}} \left( a_{\vec{R}}^{\dagger} a_{\vec{R}} + \proj{e_{\vec{R}}} \right)
\end{equation}
for Jaynes-Cummings interaction based models (\ref{JC_eff_mb}) and
\begin{equation}
 {\cal N} = \sum_{\vec{R}} \left( a_{\vec{R}}^{\dagger} a_{\vec{R}} + \proj{2_{\vec{R}}} + \proj{3_{\vec{R}}} + 2 \proj{4_{\vec{R}}} \right)
\end{equation}
for EIT based models (\ref{polariton_raw}), which lead to polaritonic Bose-Hubbard models. Here, an excitation of the atomic level 4 is equivalent to two polariton excitations since an atom needs to absorb two cavity photons to make a transition from level 1 to level 4, whereas the transition from level 1 to 2 or 3 only requires one cavity photon. The occupations of level 4 thus carry a double weight.

We note that the chemical potential $\mu$ introduced in equation (\ref{H_GC}) is conceptually different from the chemical potential $\varepsilon$ in equation (\ref{Heffect}). The approach in section \ref{sec:pbh} considers a closed system without particle exchange with the environment and $\varepsilon$ is an energy shift that affects the polaritons. The grand canonical approach in this section, in contrast, considers a situation in which particle exchange with the surrounding is permitted and is used because of its convenience for determining the phase diagrams.

The total number of excitations is conserved for both models,
$[ H, {\cal N} ] = 0$. The boundary between Mott insulator and superfluid phases
is determined by the value of $\mu$ for which adding or removing a particle does not require energy.

\subsection{Phase diagrams for one dimensional models}

The phase diagrams for the one dimensional versions of both, polaritonic Bose-Hubbard and
Jaynes-Cummings interaction based models, have been obtained by DMRG calculations (see \cite{Scholl} for a review on DMRG).
These results were presented in \cite{RF07}, where in addition the existence of a polaritonic glassy phase in setups with disorder has been predicted.

\begin{figure}
  \centering
    \includegraphics[width=\linewidth]{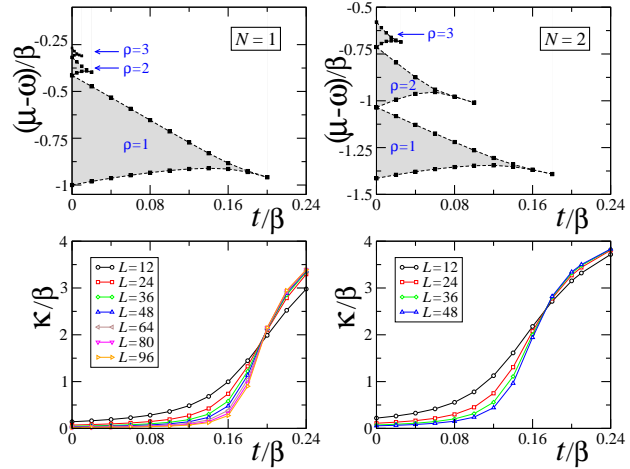}
    \caption{Upper panels: Phase diagram for the
      Hamiltonian model (\ref{JC_eff_mb}), with $N=1,2$ atoms inside each cavity
      at $\epsilon = \omega$.
      Lower panels: System compressibility $\kappa$ for the first
      lobe (i.e., $\rho = 1$), for different system sizes $L$,
      with $N=1$ (left) and $N=2$ (right). Reprinted with permission from \cite{RF07}.}
    \label{fig:PhaseDiag_I}
\end{figure}

\begin{figure}
  \centering
    \includegraphics[width=\linewidth]{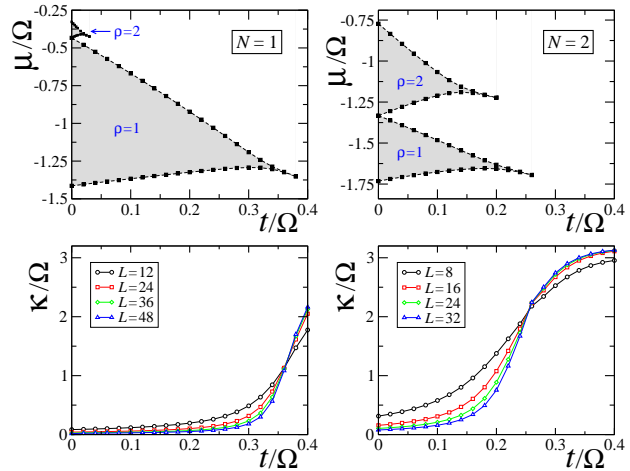}
    \caption{Upper panels: Phase diagram of model (\ref{polariton_raw}).
      The detuning parameters are set to zero and $g / \Omega = 1$.
      Lower panels: System compressibility $\kappa$  for the first
      lobe, for different system sizes $L$, with $N=1$ (left) and $N=2$ (right).
      Reprinted with permission from \cite{RF07}.}
    \label{fig:PhaseDiag_II}
\end{figure}

The authors consider a linear array of $L$ cavities with
various numbers of atoms, $N$, in each cavity. Figure \ref{fig:PhaseDiag_I} shows the phase diagram
and compressibility for the model (\ref{JC_eff_mb}). Here $\omega$ denotes the frequency of the cavity-photons ($\omega \equiv \omega_C$), $\epsilon$ the energy of the excited atomic levels
($\epsilon \equiv \omega_0$), $\beta$ the atom-photon coupling ($\beta \equiv g$) and $\rho$ the number of excitations per cavity.
The compressibility $\kappa$ is defined by $\kappa = \partial \rho / \partial \mu$.
The Mott insulator is characterised by a vanishing compressibility whereas $\kappa$ is finite for
a superfluid.
Figure \ref{fig:PhaseDiag_II} shows the phase diagram and compressibility for the model (\ref{polariton_raw}).

For a Bose-Hubbard Hamiltonian, the lobe width at $t=0$ is given by the on-site repulsion
$U$ and the critical value $t^*$ of the hopping strength at which the various lobes shrink
into a point is proportional to $U$. In contrast to the Bose-Hubbard model, in both models, (\ref{JC_eff_mb}) and (\ref{polariton_raw}), for fixed $N$, the critical values
$t^*$ are not proportional to the lobe width at $t=0$. Furthermore, the ratio between the
upper and the lower slopes of the lobes at small hopping is greater than the one
predicted for the Bose-Hubbard model ($2 n_{\text{polariton}}$ for the lower and $- 2 ( n_{\text{polariton}} + 1)$ for the upper slope \cite{FM95}); this discrepancy disappears on increasing
the number of atoms inside the cavity.
In terms of an effective Bose-Hubbard model, this may be interpreted as a
correlated hopping of the polaritons, where the hopping depends
on the occupation of the cavity. For larger occupations, the polaritons have a stronger
tendency to tunnel between cavities and consequently the superfluid regime is already reached
for lower values of the hopping term prefactor $t$. Deviations of the critical hopping $t^*$ from the critical hopping in a Bose-Hubbard model vanish for $N \gtrsim 10$  atoms per cavity \cite{RF07}.

For the detection of the Mott insulator-superfluid quantum phase transition, the
polariton and photon fluctuations in a block of linear dimension $M$
(in units of the lattice constant of the array) have been studied \cite{RFS08}.
Whereas polariton fluctuations are independent of the block size $M$ in the Mott insulator regime, they grow logarithmically with $M$ in the superfluid phase. The prefactor of the logarithm is thereby related to the compressibility of the system. For the photon fluctuations, the scaling of the leading terms is always linear in $M$ and non-critical while the critical behaviour is encoded in the subleading scaling \cite{RFS08}.

\subsection{Phase diagrams for two dimensional models}

The phase diagrams for the two dimensional version of the Jaynes-Cummings based model (\ref{JC_eff_mb}) have been calculated by various approaches, among those a mean field
approach \cite{GTCH06} and a variational cluster approach \cite{AHTL08}.

\subsubsection{Mean-field phase diagram for the 2d Jaynes-Cummings based model}

A mean-field phase diagram for the two dimensional version of the model (\ref{JC_eff_mb}) on
a triangular lattice has been obtained in \cite{GTCH06}.
Introducing the superfluid order parameter
$\Psi = \langle a_{\vec{R}} \rangle$ and using the decoupling approximation
$a_{\vec{R}}^{\dagger} a_{\vec{R}'} = \langle a_{\vec{R}}^{\dagger} \rangle a_{\vec{R}'} + a_{\vec{R}}^{\dagger} \langle a_{\vec{R}'} \rangle +
\langle a_{\vec{R}}^{\dagger} \rangle \langle a_{\vec{R}'} \rangle$, one obtains the mean-field
Hamiltonian,
\begin{align} \label{JC_eff_mb_mf}
 H^{MF} = \sum_{\vec{R}} & \left[ H^{JC}_{\vec{R}}
- z J \Psi \left(a_{\vec{R}}^{\dagger} + a_{\vec{R}}\right) + z J |\Psi|^2 \right. \nonumber \\
- & \left. \mu \left( a_{\vec{R}}^{\dagger} a_{\vec{R}} + \proj{e_{\vec{R}}}\right) \right],
\end{align}
where $z$ counts the number of nearest neighbours and is here set to $z = 3$ since a triangular lattice is assumed \cite{GTCH06}.
Here again, a chemical potential $\mu$ has been introduced in (\ref{JC_eff_mb_mf}) and the
phase boundary between superfluid and Mott insulator is given by the value of $\mu$, for which adding or removing one excitation to or from the system does not cost energy.
\begin{figure}
 \centering
 \includegraphics[width=\linewidth]{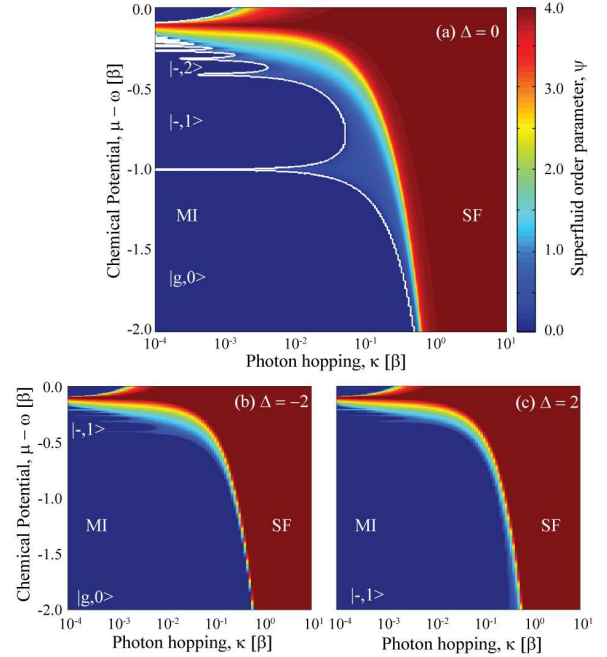}
 \caption{Slices showing the superfluid order parameter, $\Psi$, as a function of the photon hopping frequency, here denoted by $\kappa$ ($\kappa \equiv J$), and the chemical potential, $\mu$. For detuning, $\Delta=0$ ({\bf a}), $\Delta = 2 \beta$ ($\beta$ denotes the Jaynes-Cummings coupling, $\beta \equiv g$) ({\bf b}) and $\Delta = -2 \beta$ ({\bf c}) (Note that the figure defines $\Delta = \omega_C- \omega_0$ and denotes the cavity resonance by $\omega$, contrary to the text). Mott insulator lobes are indicated by the regions of $\Psi = 0$, the lowest of which are shown. Dominating the left-hand edge (where photonic repulsion dominates over hopping) is the Mott insulator phase (denoted MI), and the superfluid phase is found on the right-hand edge (denoted SF). The white contour in {\bf a} corresponds to the region where $\Psi$ becomes non-zero, delineating the quantum phase transition. Reprinted with permission from \cite{GTCH06}.} \label{meanfield_phdiag} 
\end{figure}

\begin{figure}
 \centering
 \includegraphics[width=.7\linewidth]{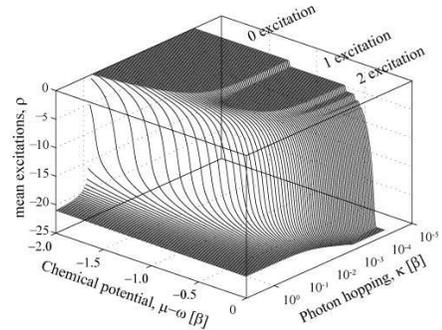}
 \caption{Plateaus with constant density, $\rho$, indicating regions with a definite
          number state excitations, as a function of the chemical potential $\mu$ and photon
          hopping frequency $\kappa$ for detuning $\Delta = 0$. Regions with varying $\rho$
          have coherent states as the ground state configuration.
          Reprinted with permission from \cite{GTCH06}.}
          \label{exct_GTCH06} 
\end{figure}

The mean-field phase diagram for the two dimensional model (\ref{JC_eff_mb_mf}) on a triangular
lattice is shown in figure \ref{meanfield_phdiag}.
Here, $\Psi = 0$ corresponds to stable Mott lobes,
where the number of excitations in each cavity increases with $\mu$.
The regions on the right correspond to $\Psi \not= 0$, where the system will be found in a
superfluid phase, i.e. within the mean field approximation, the ground state corresponds to a coherent state of photons in each cavity. The number of photons in each Mott lobe can be confirmed by considering the average number of excitations per site in the grand canonical ensemble
$\rho = \partial E_g / \partial \mu$, where $E_g$ is the ground state energy. Figure \ref{exct_GTCH06} shows $\rho$ as a function of the hopping and chemical potential. The plateaus indicate regions of constant density and incompressibility which are characteristic for
a Mott insulator phase.

A mean-field phase diagram for settings with more than one two level system in each cavity
has been calculated in \cite{Na08}, where 8 two level systems par cavity have been considered.
Such a model describes photonic crystal micro-cavities doped with substitutional donor or
acceptor impurities.

\subsubsection{Phase diagrams for 2d models from a variational cluster approach}

\begin{figure}
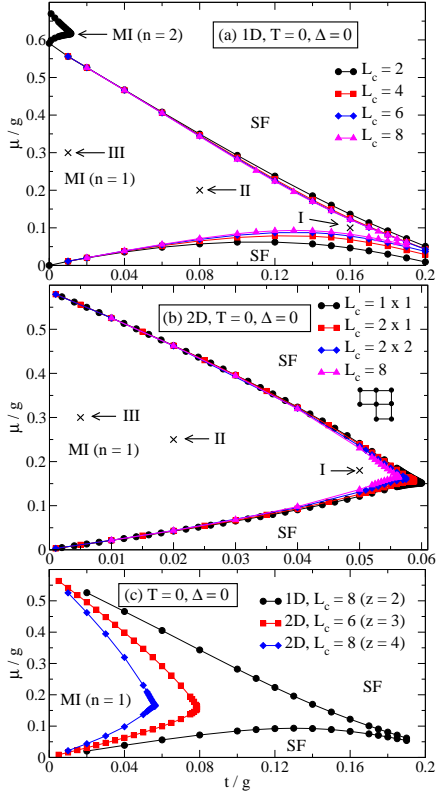

  \centering
  \includegraphics[width=0.7\linewidth]{mottlobes_T0_1D_AHTL08.eps}
  \includegraphics[width=0.7\linewidth]{mottlobes_T0_2D_AHTL08.eps}
  \includegraphics[width=0.7\linewidth]{mottlobes_T0_coord_AHTL08.eps}
  \caption{Ground-state phase diagram for the Hamiltonian (\ref{JC_eff_mb}): (a) 1D, (b) 
    2D and (c) different coordination numbers $z$. 
    Lines are guides to the eye. Reprinted with permission from \cite{AHTL08}.}
  \label{fig:pdT0_AHTL08}
\end{figure}

\begin{figure}
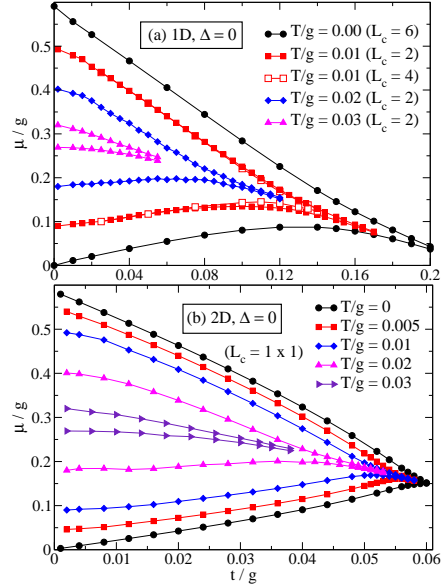

  \centering
  \includegraphics[width=0.7\linewidth]{mottlobes_Tvar_1D_AHTL08.eps}
  \includegraphics[width=0.7\linewidth]{mottlobes_Tvar_2D_AHTL08.eps}
  \caption{Mott-like regions for different $T$ for the Hamiltonian (\ref{JC_eff_mb}). Mott lobes shrink with increasing temperature $T$. Reprinted with permission from \cite{AHTL08}.}
  \label{fig:pdT>0_AHTL08}
\end{figure}

The phase diagram of model (\ref{JC_eff_mb}) for the one and two dimensional case and the
stability of its Mott insulator states with respect to non-zero temperatures have been analysed in \cite{AHTL08}. The authors employ a variational cluster approach that is expected to become exact with increasing cluster size $L_c$.
Figure \ref{fig:pdT0_AHTL08} shows the phase diagram in one dimension, two dimensions and for different
coordination numbers $z$. $\mu$ is the chemical potential, $g$ the Jaynes-Cummings coupling and
$t$ the tunnelling rate of photons between neighbouring cavities.
Figure \ref{fig:pdT>0_AHTL08} shows the unit filling Mott lobes for both, the one and two dimensional model (\ref{JC_eff_mb}) at finite temperatures $T$.

Phase diagrams for the model (\ref{JC_eff_mb}) in two dimensions have also been obtained by the stochastic series expansion quantum Monte Carlo method \cite{ZSU08}, where it has been predicted that the universality class of coupled cavity arrays may be different from the Bose-Hubbard model.

\subsection{Bose-glass phases}

\begin{figure}
  \centering
    \includegraphics[width=0.8\linewidth]{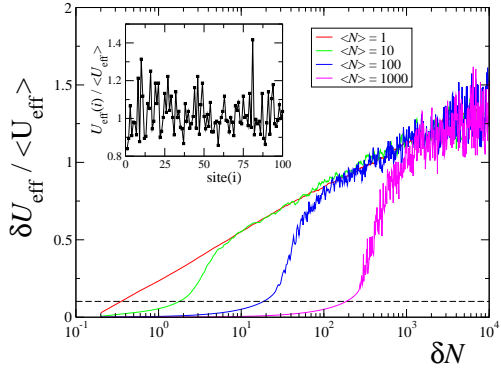}
    \caption{Relative standard deviation of the effective
      on-site interaction strength $U_{\rm eff}$ averaged over $L=10^4$
      cavities, as a function of the atom number fluctuations $\delta N$; 
      from left to right $\left< N \right> = 1, \, 10, \, 100, \, 1000$.
      Dashed line indicates an effective variation of the interaction strength
      equal to the standard deviation of a random interaction uniformly
      distributed in the interval
      $U_{\rm eff} (i) / \left< U_{\rm eff} \right> \in [-\epsilon,+\epsilon]$
      with $\epsilon=0.25$.
      In the inset an example of the variation on the on-site effective
      interaction is shown; $\left< N \right> = 100$, $\delta N = 20$.
      Reprinted with permission from \cite{RF07}.}
    \label{fig:Glass_I}
\end{figure}

A glassy phase for polaritons in a one dimensional cavity array with disorder has been predicted in \cite{RF07}.
This phase is characterised by a finite compressibility, gapless excitation spectrum and zero superfluid density and has been predicted to exist in a Bose-Hubbard model for relative fluctuations
of the interaction, $\delta U_{\rm eff} / \left< U_{\rm eff} \right>$, uniformly
distributed in an interval of length $2\epsilon = 0.5$ in a system of $L=200$ sites at filling
$\rho =1.01$ for $0.078 \lesssim t_{\rm eff} / \left< U_{\rm eff} \right> \lesssim 0.133$.
Figure \ref{fig:Glass_I} shows relative fluctuations of the interaction $\delta U_{\rm eff} / \left< U_{\rm eff} \right>$ as a function of atom number fluctuations over an interval $\delta N$
in the cavities. For the model (\ref{JC_eff_mb}), a glassy phase should be visible for
, e.g. $\left< N \right> =100$ particles per cavity and $\delta N \simeq 19$ (such to have $\epsilon = 0.25$) in the range
$7.51 \times 10^{-3} \lesssim t/\beta \lesssim 1.401 \times 10^{-2} \;.$

The observability of quantum phase transitions from a superfluid to a Bose-glass to a Mott insulator that is driven by an decrease of the photon tunnelling rate $J$ has also been predicted for a two dimensional model in \cite{Na08}. Here, photonic crystal micro-cavities doped with substitutional donor or acceptor impurities have been considered, where fluctuations in the number of impurities ($\Delta N$), the impurity-photon coupling ($\Delta g_k$) and the detunings between impurity and photon resonances ($\Delta \omega_{\text{ph}}$) are expected in experimental realisations.
\begin{figure}
  \centering
    \includegraphics[width=0.8\linewidth]{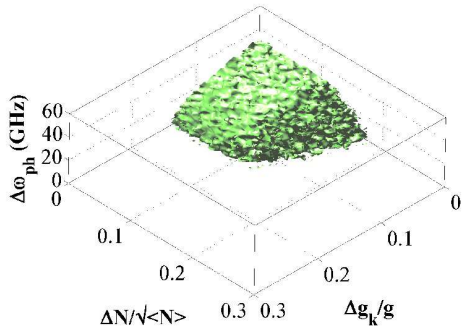}
    \caption{The isovalue surface that separates the
existence and nonexistence of an equilibrium polaritonic quantum phase transition from
the Bose glass to Mott insulator states. Reprinted with permission from \cite{Na08}.}
    \label{fig:Glass_II}
\end{figure}
The information whether an equilibrium polaritonic quantum phase transition from
a Bose glass to Mott insulator states exists is obtained by
plotting the isosurface in which the critical polariton tunnelling
rate at the Bose glass Mott insulator phase boundary is set to be 10 times
larger than the polariton loss rate, assuming a cavity Q-factor $10^6$, an impurity number $N=3$ and an impurity photon detuning $\Delta = 12 g$.
Inside the isovalue surface, quantum phase transitions from a superfluid to a
Bose glass to Mott insulator states can be observed by continuously decreasing the tunnelling rate $J$.
Outside the isovalue surface, only superfluid and Bose glass phases exist.


\section{Effective Spin Models in Coupled Cavity Arrays} \label{sec:spin}

In the previous sections we have seen that under appropriate conditions interesting bosonic many-body Hamiltonians can be created and probed in coupled cavity arrays. There the atoms were used for detection and most importantly to generate an interaction between photons in the same cavity, either actively, as in the polaritonic case, or in an indirect manner, as for the creation of large Kerr nonlinearities. In this section we focus on a complementary regime, in which the tunnelling of photons is used to mediate interactions for atoms in neighbouring cavities. We review proposals to realise effective spin lattice Hamiltonians with individual atoms in micro-cavities that are coupled to each other \cite{HBP07a,KA08,CAB08}.

Interacting spin systems have a key role in quantum statistical mechanics, condensed-matter physics, and more recently in quantum information science. On the one hand they have a simple microscopic description, which allows the use of techniques from statistical mechanics in their analysis. On the other, they have a very rich structure \cite{MU03} and can hence be employed to model and learn about phenomena of real condensed-matter systems. In quantum information science, spin chains can be harnessed to propagate and manipulate quantum information, as well as to implement quantum computation. We first discuss how spin models that conserve the total number of excitations can be generated as limiting cases of Bose-Hubbard type models. Then we review an approach where the full anisotropic Heisenberg Hamiltonian can be engineered for atoms that interact
via the exchange of virtual photons. Finally we mention schemes that investigate magnetic properties for photons of different polarisations.

\subsection{Spin models that conserve the total number of excitations} \label{sec:JCspin}

A certain class of effective spin Hamiltonians can be generated as a limiting case of the
effective many-body Hamiltonians discussed in sections \ref{BHmodel} and \ref{sec:JC}.
Whenever the engineered on-site interactions are repulsive and strong enough to suppress the
occupation of states with more than one excitation almost completely, the states with zero and one
excitation can be identified with $\ket{\uparrow}$ and $\ket{\downarrow}$ respectively \cite{ASB07} and the dynamics can effectively be described by the Hamiltonian
\begin{equation} \label{XY_cons}
H = \sum_j B \sigma_j^z + \sum_{<j, j'>} J ( \sigma_j^x \sigma_{j'}^x +  \sigma_j^y \sigma_{j'}^y ),
\end{equation}
where the second sum runs over nearest neighbours. $\sigma_j^{x/y/z}$ are the Pauli matrices at site $j$, $J$ a coupling coefficient and $B$ a magnetic field. 

The Hamiltonian (\ref{XY_cons}) conserves the number of spins in the state $\ket{\uparrow}$ and
can thus be solved efficiently. Much richer spin Hamiltonians can be generated when putting
atoms of a $\Lambda$ level structure into the cavities.

\subsection{The anisotropic Heisenberg model}\label{Heisenberg}

A particular interesting spin Hamiltonian is the anisotropic Heisenberg or XYZ model, given by
\begin{align} \label{XYZ}
H = & \sum_j B \sigma_j^z \\
+ & \sum_{<j, j'>} J_x \sigma_j^x \sigma_{j'}^x + J_y \sigma_j^y \sigma_{j'}^y + J_z \sigma_j^z \sigma_{j'}^z \nonumber .
\end{align}
This model contains a rich phase diagram and can be used to study quantum magnetism as a limiting case of the fermionic Hubbard model, which is believed to contains the main features of high $T_c$ superconductors \cite{MU03}. From a quantum information perspective, such a model can be used to create cluster states, highly entangled multipartite quantum states which are universal for quantum computation by single-site measurements \cite{RB01}.

The Hamiltonian (\ref{XYZ}) can be generated in an array of coupled cavities where each cavity contains one atom of a $\Lambda$ level structure, c.f. figure \ref{xy_levels} and \ref{zz_levels}. The two spin polarisations $\ket{\uparrow}$ and $\ket{\downarrow}$ are represented by the two longlived atomic levels of the $\Lambda$ level structures. Together with external lasers, the cavity mode that couples to the atom inside each cavity can induce Raman transitions between these two longlived levels. Due to a large enough detuning between laser and cavity mode, these transitions can only create virtual photons in the cavity mode which mediate an interaction with another atom in a neighbouring cavity. With appropriately chosen detunings, both the excited atomic levels and photon states have vanishing occupation and can be eliminated from the description. As a result, the dynamics is confined to only two states per atom, the long-lived levels, and can be 
described by a spin 1/2 Hamiltonian. The small occupation of photon states and excited atomic levels also strongly suppresses spontaneous emission and cavity decay. As discussed later, this proposal can be realised as soon as the strong coupling regime of cavity QED is achieved.

We note that similar ideas have been recently proposed to realise spin systems of larger spins \cite{CAB08} and with interactions that depend on the direction of the respective lattice edge \cite{KA08}.

In the next subsections we first discuss how effective $\sigma^x \sigma^x$, $\sigma^y \sigma^y$ and $\sigma^z \sigma^z$ interactions together with an effective magnetic field $B \sigma_z$ can be engineered and then explain how these can be combined to generate the full anisotropic Heisenberg 
model (\ref{XYZ}).

\subsubsection{XX and YY interactions}

\begin{figure}
\centering
\psfrag{a}{\raisebox{-0.08cm}{$a$}}
\psfrag{b}{\raisebox{-0.08cm}{$b$}}
\psfrag{e}{$e$}
\psfrag{da}{\raisebox{-0.06cm}{$\delta_a$}}
\psfrag{db}{\raisebox{-0.06cm}{$\delta_b$}}
\psfrag{Oa}{\hspace{-0.14cm}$\Omega_a$}
\psfrag{Ob}{\raisebox{-0.04cm}{\hspace{-0.06cm}$\Omega_b$}}
\psfrag{ga}{\hspace{-0.04cm}$g_a$}
\psfrag{gb}{$g_b$}
\psfrag{Da}{\hspace{-0.14cm}$\Delta_a$}
\psfrag{Db}{\hspace{-0.08cm}$\Delta_b$}
\psfrag{wa}{\hspace{-0.1cm}$\omega_a$}
\psfrag{wb}{\hspace{-0.04cm}$\omega_b$}
\psfrag{wab}{\hspace{-0.04cm}$\omega_{ab}$}
\includegraphics[width=.9\linewidth]{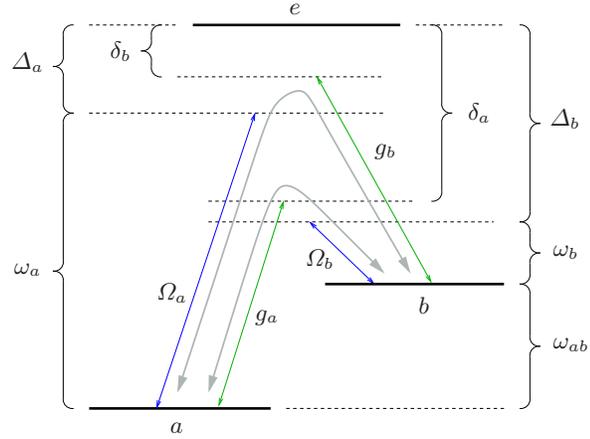}
\caption{\label{xy_levels} Level structure, driving lasers and relevant couplings to the cavity mode to generate effective $\sigma^x \sigma^x$- and $\sigma^y \sigma^y$-couplings for one atom. The cavity mode couples with strengths $g_a$ and $g_b$ to transitions $\ket{a} \leftrightarrow \ket{e}$ and $\ket{b} \leftrightarrow \ket{e}$ respectively. One laser with frequency $\omega_a$ couples to transition $\ket{a} \leftrightarrow \ket{e}$ with Rabi frequency $\Omega_a$ and another laser with frequency $\omega_b$ to $\ket{b} \leftrightarrow \ket{e}$ with $\Omega_b$. The dominant 2-photon processes are indicated in faint grey arrows. Reprinted with permission from \cite{HBP07a}.}
\end{figure}
XX and YY interactions can be generated in an array of coupled cavities with one 3-level atom in each cavity (see figure \ref{xy_levels}). The two long lived levels, $\ket{a}$ and $\ket{b}$, represent the two spin states. The cavity mode couples to the transitions 
$\ket{a} \leftrightarrow \ket{e}$ and $\ket{b} \leftrightarrow \ket{e}$, 
where $\ket{e}$ is the excited state of the atom. Furthermore, 
two driving lasers couple to the transitions 
$\ket{a} \leftrightarrow \ket{e}$ respectively $\ket{b} \leftrightarrow \ket{e}$.
For the sake of simplicity we consider here a one-dimensional array, but the generalisation to higher dimensions is straight forward. The Hamiltonian of the atoms reads
\begin{equation}
H_A = \sum_{j=1}^N \omega_e \ket{e_j} \bra{e_j} + \omega_{ab} \ket{b_j} \bra{b_j}, 
\end{equation}
where the index $j$ counts the cavities, $\omega_e$ is the 
energy of the excited level and $\omega_{ab}$ the energy of 
level $\ket{b}$. The energy of level $\ket{a}$ is set to zero. 
The Hamiltonian that describes the photons in the cavity modes is 
given by equation. (\ref{arrayham2}). For simplicity of notation we set
$J_C = 2 \omega_C \alpha$ and for convenience we assume periodic boundary 
conditions, where $H_C$ can be diagonalised via the Fourier transform
\begin{equation*}
a_k = \frac{1}{\sqrt{N}} \sum_{j=1}^N \ee^{\iu k j} a_j \, , \: \:
k = \frac{2 \pi l}{N} \, , \: \: - \frac{N}{2} \le l \le \frac{N}{2}
\end{equation*}
for $N$ odd to give $H_C = \sum_{k} \omega_k a_k^{\dagger} a_k$ with $\omega_k = \omega_C + 2 J_C \cos(k)$. Finally the interaction between the atoms and the photons 
as well as the driving by the lasers are described by
\begin{align}
H_{AC} = \sum_{j=1}^N & \left[ \left(\frac{\Omega_a}{2} \ee^{-\iu \omega_a t} + g_a a_j \right) \ket{e_j} \bra{a_j} \right. \\
+ & \left. \left(\frac{\Omega_b}{2} \ee^{-\iu \omega_b t} + g_b a_j \right) \ket{e_j} \bra{b_j} + \text{h.c.} \right] \nonumber .
\end{align}
Here $g_a$ and $g_b$ are the couplings of the respective transitions to the cavity mode, $\Omega_a$ is the Rabi frequency of one laser with frequency $\omega_a$ and  $\Omega_b$
the Rabi frequency of a second laser with frequency $\omega_b$ \cite{SM02}. The complete Hamiltonian is then given by $H = H_A + H_C + H_{AC}$ .

By switching to an interaction picture with respect to $H_0 = H_A + H_C - \delta_1 \, \sum_{j=1}^N \ket{b_j} \bra{b_j}$, where $\delta_1 = \omega_{ab} - (\omega_a - \omega_b)/2$, the excited atom levels $\ket{e_j}$ and the photons can be adiabatically eliminated \cite{J00}. To obtain effective spin-spin interactions, terms up to 2nd order need to be considered while dropping fast oscillating terms. For this approach the detunings $\Delta_a \equiv \omega_e - \omega_a$, $\Delta_b \equiv \omega_e - \omega_b - (\omega_{ab} - \delta_1)$, $\delta_a^k \equiv \omega_e - \omega_k$ and $\delta_b^k \equiv \omega_e - \omega_k - (\omega_{ab} - \delta_1)$ have to be large compared to the couplings $\Omega_a, \Omega_b, g_a$ and $g_b$. Furthermore, the parameters must be such that the dominant Raman transitions between levels $a$ and $b$ are those that involve one laser photon and one cavity photon each (c.f. figure \ref{xy_levels}). To avoid excitations of real photons via these transitions one requires
$\left| \Delta_a - \delta_b^k \right|, \left| \Delta_b - \delta_a^k \right| \gg \left| \frac{\Omega_a g_b}{2 \Delta_a} \right|, \left| \frac{\Omega_b g_a}{2 \Delta_b} \right|$.
 
Hence whenever the atom emits or absorbs a virtual photon into or from the cavity mode, it does a transition from level $\ket{a}$ to $\ket{b}$ or vice versa. If one atom emits a virtual photon in such a process that is absorbed by a neighbouring atom, which then also does a transition between 
$\ket{a}$ to $\ket{b}$, an effective spin-spin interaction has happened. Dropping irrelevant constants, the resulting effective Hamiltonian reads
\begin{equation}
H_{\text{xy}} = \sum_{j=1}^N B \sigma_j^z
+ ( J_1 \sigma_j^+ \sigma_{j+1}^- + J_2 \sigma_j^- \sigma_{j+1}^- + \text{h.c.} ),
\end{equation}
where $\sigma_j^z = \ket{b_j} \bra{b_j} - \ket{a_j}\bra{a_j}$, $\sigma_j^+ = \ket{b_j}\bra{a_j}$ and $B$, 
$J_1$ and $J_2$ are functions of the various Rabi frequencies and detunings\footnote{To second order they are given by
$B = \frac{\delta_1}{2} - \frac{1}{2} \left[\frac{|\Omega_b|^2}{4 \Delta_b^2} 
\left( \Delta_b - \frac{|\Omega_b|^2}{4 \Delta_b} - \frac{|\Omega_a|^2}{4 (\Delta_a - \Delta_b)} - \gamma_b g_b^2 - \gamma_1 g_a^2 \right. \right.$\\
$ \left. \left. + \gamma_1^2 \frac{g_a^4}{\Delta_b} \right) -
(a \leftrightarrow b) \right]$, 
$J_1 = \frac{\gamma_2}{4} \left( \frac{|\Omega_a|^2 g_b^2}{\Delta_a^2} + \frac{|\Omega_b|^2 g_a^2}{\Delta_b^2} \right)$
and
$J_2 = \frac{\gamma_2}{2} \frac{\Omega_a^{\star} \Omega_b g_a g_b}{\Delta_a \Delta_b}$,
where
$\gamma_{a,b} = \frac{1}{N} \sum_{k} \frac{1}{\omega_{a,b} -\omega_k}$,
$\gamma_1 = \frac{1}{N} \sum_{k} \frac{1}{(\omega_a + \omega_b)/2 - \omega_k}$ and
$\gamma_2 = \frac{1}{N} \sum_{k} \frac{\exp (\iu k)}{(\omega_a + \omega_b)/2 - \omega_k}$.
}. If $J_2^{\star} = J_2$, this Hamiltonian reduces to the XY model,
\begin{equation} \label{HXYb}
H_{\text{xy}} = \sum_{j=1}^N B \sigma_j^z + J_x \sigma_j^x 
\sigma_{j+1}^x + J_y \sigma_j^y \sigma_{j+1}^y \, ,
\end{equation}
with $J_x = (J_1 + J_2)/2$ and $J_y = (J_1 - J_2)/2$.

For $\Omega_a = \pm (\Delta_a g_a / \Delta_b g_b) \Omega_b$ with $\Omega_a$ and $\Omega_b$ real, the interaction is either purely $\sigma^x \sigma^x$ ($+$) or purely $\sigma^y \sigma^y$ ($-$) and the Hamiltonian (\ref{HXYb}) becomes the Ising model in a transverse field, whereas the isotropic XY model ($J_x = J_y$, i.e. $J_2 = 0$) is obtained for either $\Omega_a \rightarrow 0$ or $\Omega_b \rightarrow 0$. The effective magnetic field $B$ in turn can, independently of $J_x$ and $J_y$, be tuned to assume any value between $|B| \gg |J_x|, |J_y|$ and $|B| \ll |J_x|, |J_y|$ by varying $\delta_1$. Hence the system can be driven through quantum phase transition by varying the Rabi frequencies and detunings of the driving lasers. Now we proceed to discuss effective ZZ interactions.

\subsubsection{ZZ interactions}

\begin{figure}
\centering
\psfrag{a}{\raisebox{-0.04cm}{$a$}}
\psfrag{b}{\raisebox{-0.04cm}{$b$}}
\psfrag{e}{$e$}
\psfrag{da}{\raisebox{-0.12cm}{\hspace{0.4cm} $\tilde{\Delta}_a$}}
\psfrag{-db}{\hspace{0.14cm} $-\tilde{\Delta}_b$}
\psfrag{ka}{\hspace{-0.1cm}$\delta_a$}
\psfrag{kb}{\raisebox{-0.04cm}{\hspace{-0.1cm}$\delta_b$}}
\psfrag{Oa}{\hspace{-0.1cm}$\Omega_a$}
\psfrag{Ob}{\hspace{-0.06cm}$\Omega_b$}
\psfrag{La}{$\Lambda_a$}
\psfrag{Lb}{$\Lambda_b$}
\psfrag{ga}{\hspace{-0.06cm}$g_a$}
\psfrag{gb}{\hspace{-0.1cm}$g_b$}
\psfrag{Da}{\hspace{-0.1cm}$\Delta_a$}
\psfrag{Db}{$\Delta_b$}
\psfrag{n}{$\nu$}
\psfrag{w}{$\omega$}
\psfrag{wb}{$\omega_b$}
\psfrag{wab}{$\omega_{ab}$}
\includegraphics[width=.9\linewidth]{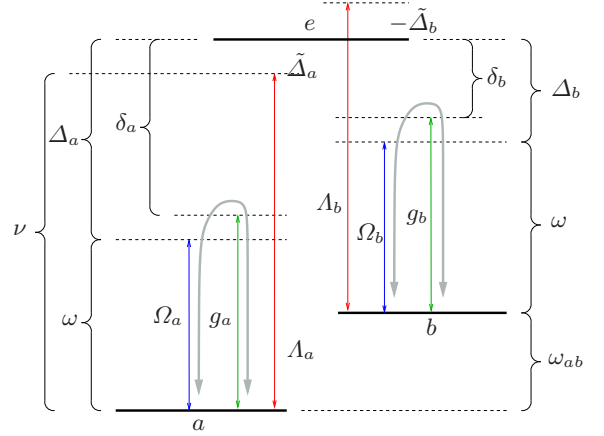}
\caption{\label{zz_levels} Level structure, driving lasers and relevant couplings to the cavity mode to generate effective $\sigma^z \sigma^z$-couplings for one atom. The cavity mode couples with strengths $g_a$
and $g_b$ to transitions $\ket{a} \leftrightarrow \ket{e}$ and $\ket{b} \leftrightarrow \ket{e}$ respectively. Two lasers with frequencies $\omega$ and $\nu$ couple with Rabi frequencies $\Omega_a$ respectively $\Lambda_a$ to transition $\ket{a} \leftrightarrow \ket{e}$ and $\Omega_b$ respectively $\Lambda_b$ to $\ket{b} \leftrightarrow \ket{e}$. The dominant 2-photon processes are indicated in faint grey arrows. Reprinted with permission from \cite{HBP07a}.}
\end{figure}

Effective $\sigma^z \sigma^z$ interactions can be generated with the same atomic level configuration as XX and YY terms, but now only one laser with frequency $\omega$ mediates atom-atom coupling via virtual photons. A second laser with frequency $\nu$ is used to tune the effective magnetic field via a Stark shift. The atoms together with their couplings to cavity mode and lasers are shown in figure \ref{zz_levels}.
Again, we consider the one-dimensional case as an example. The generalisation to higher dimensions is straightforward.
The Hamiltonians $H_A$ of the atoms and $H_C$ of the cavity modes thus have the same form as above, whereas $H_{AC}$ now reads
\begin{align}
H_{AC} = & \sum_{j=1}^N \left[ \left(\frac{\Omega_a}{2} \ee^{-\iu \omega t} +
\frac{\Lambda_a}{2} \ee^{-\iu \nu t} + g_a a_j \right) \ket{e_j}\bra{a_j} \right. \\
+ & \left. \left(\frac{\Omega_b}{2} \ee^{-\iu \omega t} +
\frac{\Lambda_b}{2} \ee^{-\iu \nu t} + g_b a_j \right) \ket{e_j}\bra{b_j} + \text{h.c.} \right] \nonumber .
\end{align}
Here, $\Omega_a$ and $\Omega_b$ are the Rabi frequencies of the driving laser with frequency $\omega$ on transitions $\ket{a} \rightarrow \ket{e}$ and $\ket{b} \rightarrow \ket{e}$, whereas $\Lambda_a$ and $\Lambda_b$ are the Rabi frequencies of the driving laser with frequency $\nu$ on transitions $\ket{a} \rightarrow \ket{e}$ and $\ket{b} \rightarrow \ket{e}$.

In an interaction picture with respect to $H_0 = H_A + H_C$, the excited atom levels $\ket{e_j}$ and the photons can be adiabatically eliminated \cite{J00}. Again, the detunings $\Delta_a \equiv \omega_e - \omega$, $\Delta_b \equiv \omega_e - \omega - \omega_{ab}$, $\tilde{\Delta}_a \equiv \omega_e - \nu$, $\tilde{\Delta}_b \equiv \omega_e - \nu - \omega_{ab}$, $\delta_a^k \equiv \omega_e - \omega_k$ and $\delta_b^k \equiv \omega_e - \omega_k - \omega_{ab}$ have to be large compared to the couplings $\Omega_a, \Omega_b, \Lambda_a, \Lambda_b, g_a$ and $g_b$, whereas now Raman transitions between levels $a$ and $b$ should be suppressed. Hence parameters must be such that the dominant 2-photon processes are those that involve one laser photon and one cavity photon each but where the atom does no transition between levels $a$ and $b$ (c.f. figure \ref{zz_levels}). To avoid excitations of real photons in these processes, one requires
$\left| \Delta_a - \delta_a^k \right|, \left| \Delta_b - \delta_b^k \right| \gg \left| \frac{\Omega_a g_a}{2 \Delta_a} \right|, \left| \frac{\Omega_b g_b}{2 \Delta_b} \right|$.

Whenever two atoms exchange a virtual photon in this scheme, both experience a Stark shift that depends on the state of the partner atom. This conditional Stark shift plays the role of an effective $\sigma^z \sigma^z$-interaction. Dropping irrelevant constants, the effective Hamiltonian reads
\begin{equation} \label{HZZ}
H_{\text{zz}} = \sum_{j=1}^N \left( \tilde{B} \sigma_j^z + J_z \sigma_j^z \sigma_{j+1}^z \right) \, .
\end{equation}
where $\tilde{B}$ and $J_z$ are determined by the detunings and Rabi frequencies of the employed lasers\footnote{To second order they are give by
$\tilde{B} = - \frac{1}{2} \left[
\frac{|\Lambda_b|^2}{16 \tilde{\Delta}_b^2} \right.$\\
$\left(4 \tilde{\Delta}_b
- \frac{|\Lambda_a|^2}{\tilde{\Delta}_a - \tilde{\Delta}_b} - 
\frac{|\Lambda_b|^2}{\tilde{\Delta}_b} - \sum_{j=a,b} \left(\frac{|\Omega_j|^2}{\Delta_j - 
\tilde{\Delta}_b} + 4 \tilde{\gamma}_{jb} g_j^2 \right) \right) +$\\
$\frac{|\Omega_b|^2}{16 \Delta_b^2}
\left(4 \Delta_b - \frac{|\Omega_a|^2}{\Delta_a - \Delta_b} - \frac{|\Omega_b|^2}{\Delta_b}
- \sum_{j=a,b} \left(\frac{|\Lambda_j|^2}{\tilde{\Delta}_j - \Delta_b} + 4 \gamma_{jb} g_j^2 \right) \right.$ \\
$ \left. \left. + 4 \gamma_{bb}^2 \frac{g_b^4}{\Delta_b} \right)
- \left( a \leftrightarrow b \right) \right]$ and
$J_z = \gamma_2 \left| \frac{\Omega_b^{\star} g_b}{4 \Delta_b} - \frac{\Omega_a^{\star} g_a}{4 \Delta_a} \right|^2$
with
$\gamma_1 = \frac{1}{N} \sum_{k} \frac{1}{\omega - \omega_k}$,
$\gamma_2 = \frac{1}{N} \sum_{k} \frac{\exp (\iu k)}{\omega - \omega_k}$,
$\gamma_{aa} = \gamma_{bb} = \frac{1}{N} \sum_{k} \frac{1}{\omega - \omega_k}$,
$\left. \begin{array}{c}
\gamma_{ab}\\
\gamma_{ba}
\end{array} \right\}= \frac{1}{N} \sum_{k} \frac{1}{\omega \pm \omega_{ab} - \omega_k}$,
$\left. \begin{array}{c}
\tilde{\gamma}_{ab}\\
\tilde{\gamma}_{ba}
\end{array} \right\}= \frac{1}{N} \sum_{k} \frac{1}{\nu \pm \omega_{ab} - \omega_k}$,
$\tilde{\gamma}_{aa} = \tilde{\gamma}_{bb} = \frac{1}{N} \sum_{k} 
\frac{1}{\nu - \omega_k}$.
}.
Here again, the interaction $J_z$ and the field $\tilde{B}$ can be tuned independently, either by varying $\Omega_a$ and $\Omega_b$ for $J_z$ or by varying $\Lambda_a$ and $\Lambda_b$ for $\tilde{B}$. In particular, $|\Lambda_a|^2$ and $|\Lambda_b|^2$ can for all values of $\Omega_a$ and $\Omega_b$ be chosen such that either $J_z \ll \tilde{B}$ or $J_z \gg \tilde{B}$.

Both, XX and YY interactions and ZZ interactions can be combined to engineer the full Hamiltonian (\ref{XYZ}). This can be done via the Suzuki-Trotter formula.

\subsubsection{The complete effective model} 

Making use of the Suzuki-Trotter formula, the two Hamiltonians (\ref{HXYb}) and (\ref{HZZ}) can now be combined to one effective Hamiltonian. To this end, the lasers that generate the Hamiltonian (\ref{HXYb}) are turned on for a short time interval $dt$ ($||H_{\text{xy}}|| \cdot dt \ll 1$) followed by another time interval $dt$ ($||H_{\text{zz}}|| \cdot dt \ll 1$) with the lasers that generate the Hamiltonian (\ref{HZZ}) turned on. This sequence is repeated until the total time 
range to be simulated is covered.

The effective Hamiltonian simulated by this procedure is $H_{\text{spin}} = H_{\text{xy}} + H_{\text{zz}}$, which is precisely the Heisenberg anisotropic model of equation. (\ref{XYZ}) with $B_{\text{tot}} = B + \tilde{B}$. The time interval $dt$ should thereby be chosen such that $\Omega^{-1}, g^{-1} \ll  dt_1 , dt_2 \ll J_x^{-1} , J_y^{-1} , J_z^{-1} , B^{-1}$ and $\tilde{B}^{-1}$, so that the Trotter sequence concatenates the effective Hamiltonians $H_{XY}$ and $H_{ZZ}$. The procedure can be generalised to higher order Trotter formulae or by turning on the sets of lasers for time intervals of different length.

\begin{figure}
\centering
\psfrag{t}{\raisebox{-0.4cm}{\scriptsize \hspace{-0.6cm} $t \: \text{in} \: 10^{-6} \: \text{s}$}}
\psfrag{pa}{\raisebox{0.4cm}{\scriptsize \hspace{-0.6cm} $p(a_1), \: p( \downarrow_1 )$}}
\psfrag{A}{\bf \hspace{-2.1cm} b}
\psfrag{B}{\bf \hspace{-2.1cm} a}
\psfrag{0a}{\raisebox{-0.12cm}{\scriptsize $0$}}
\psfrag{5a}{\raisebox{-0.12cm}{\scriptsize $5$}}
\psfrag{10a}{\raisebox{-0.12cm}{\scriptsize $10$}}
\psfrag{15a}{\raisebox{-0.12cm}{\scriptsize $15$}}
\psfrag{20a}{\raisebox{-0.12cm}{\scriptsize $20$}}
\psfrag{25a}{\raisebox{-0.12cm}{\scriptsize $25$}}
\psfrag{30a}{\raisebox{-0.12cm}{\scriptsize $30$}}
\psfrag{40a}{\raisebox{-0.12cm}{\scriptsize $40$}}
\psfrag{50a}{\raisebox{-0.12cm}{\scriptsize $50$}}
\psfrag{60a}{\raisebox{-0.12cm}{\scriptsize $60$}}
\psfrag{0}{\hspace{-0.3cm} \scriptsize $0$}
\psfrag{0.1}{\hspace{-0.4cm} \scriptsize $ $}
\psfrag{0.2}{\hspace{-0.4cm} \scriptsize $0.2$}
\psfrag{0.3}{\hspace{-0.4cm} \scriptsize $ $}
\psfrag{0.4}{\hspace{-0.4cm} \scriptsize $0.4$}
\psfrag{0.5}{\hspace{-0.4cm} \scriptsize $ $}
\psfrag{0.6}{\hspace{-0.4cm} \scriptsize $0.6$}
\psfrag{0.7}{\hspace{-0.4cm} \scriptsize $ $}
\psfrag{0.8}{\hspace{-0.4cm} \scriptsize $0.8$}
\psfrag{0.9}{\hspace{-0.4cm} \scriptsize $ $}
\psfrag{1}{\hspace{-0.3cm} \scriptsize $1$}
\hspace{0.2cm}
\includegraphics[width=.46\linewidth]{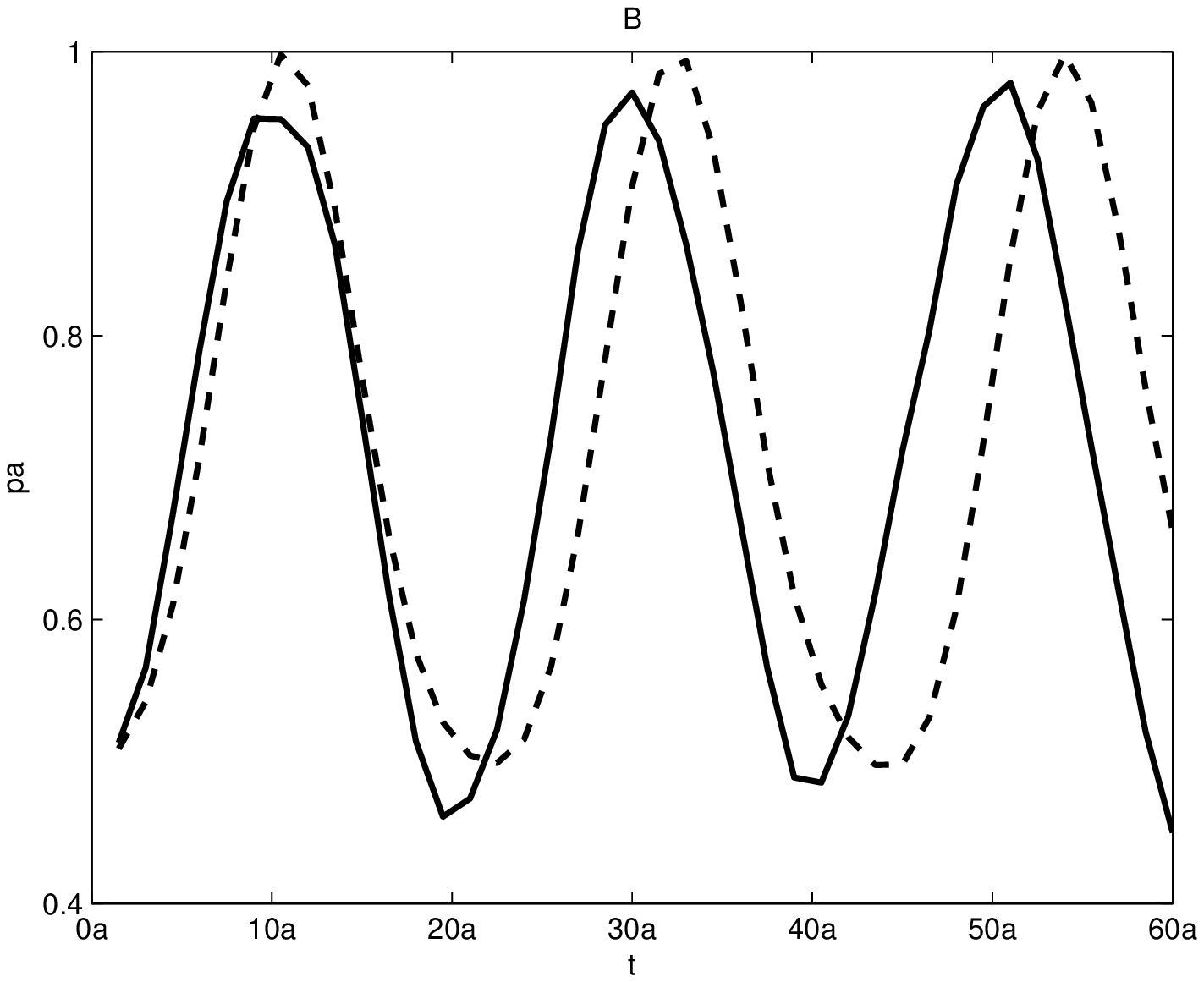}
\hfill
\includegraphics[width=.46\linewidth]{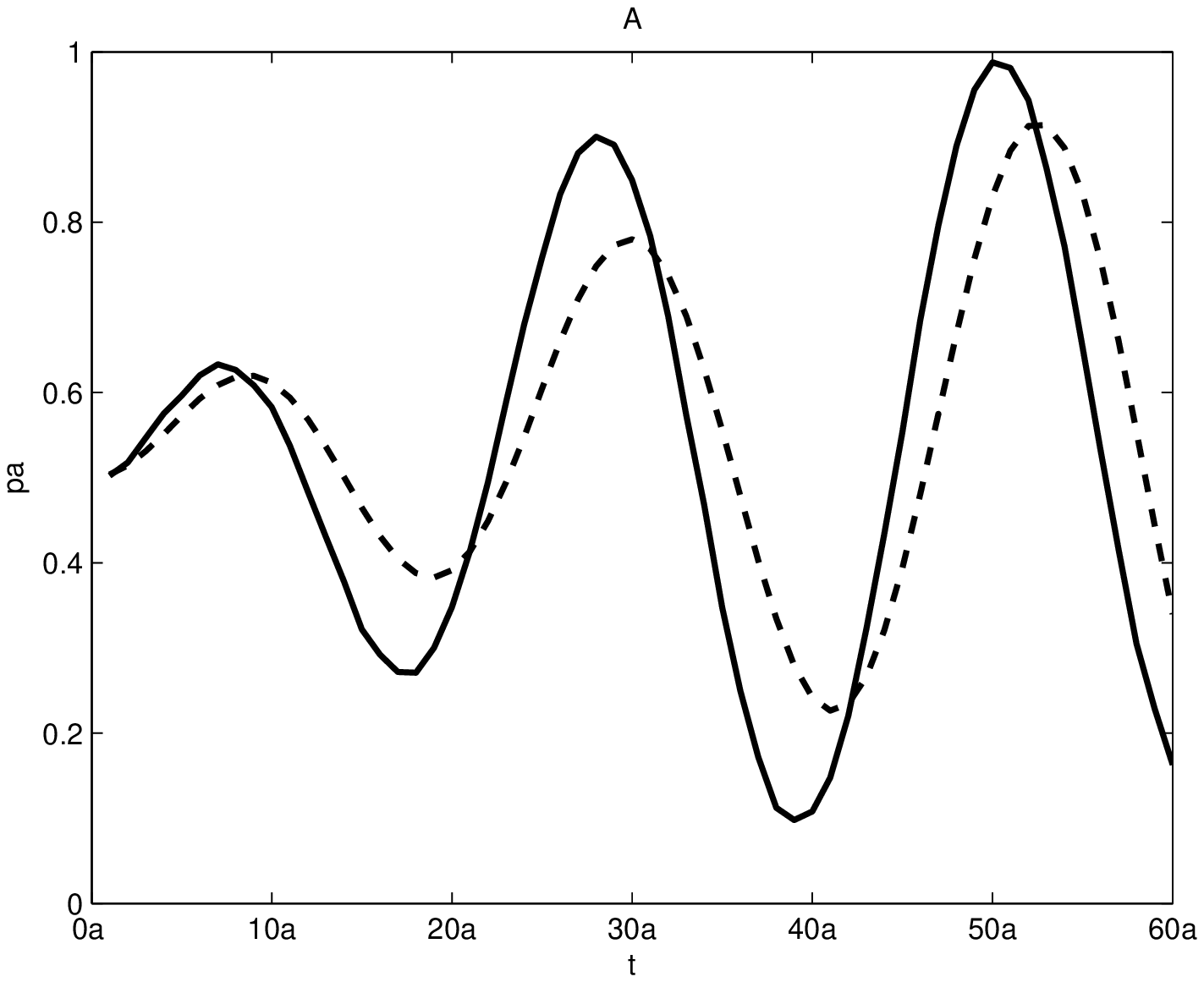}
\caption{The occupation probability $p(a_1)$ of state $\ket{a_1}$ (solid line) and the
probability $p( \downarrow_1 )$ of spin 1 to point down (dashed line) for the parameters
$\omega_e = 10^6$GHz, $\omega_{ab} = 30$GHz, 
$\Delta_a = 30$GHz, $\Delta_b = 60$GHz, $\omega_C = \omega_e - \Delta_b + 2$GHz, $\tilde{\Delta}_a = 15$GHz, $\Omega_a = \Omega_b = 2$GHz, $\Lambda_a = \Lambda_b = 0.71$GHz, $g_a = g_b = 1$GHz, $J_C = 0.2$GHz and
$\delta_1 = -0.0165$GHz (plot {\bf a}) respectively $\delta_1 = -0.0168$GHz (plot {\bf b}). Both, the occupation of the excited atomic states $\langle \ket{e_j} \bra{e_j} \rangle$ and the photon number $\langle a^{\dagger} a \rangle$ are always smaller than 0.03.
Reprinted with permission from \cite{HBP07a}.}
\label{run2}
\end{figure}

The validity of all above approximations is shown in figure \ref{run2}, where numerical simulations of the dynamics generated by the full Hamiltonian $H$ are compared it to the dynamics generated by the effective model (\ref{HXYb}). The present example considers two atoms in two cavities, initially in the state $\frac{1}{\sqrt{2}} (\ket{a_1} + \ket{b_1}) \otimes \ket{a_2}$, and calculates the occupation probability $p(a_1)$ of the state $\ket{a_1}$ which corresponds to the probability of spin 1 to point down, $p( \downarrow_1 )$. Figure \ref{run2}{\bf a} shows $p(a_1)$ and $p( \downarrow_1 )$ for an effective Hamiltonian (\ref{XYZ}) with $B_{\text{tot}} = 0.135$MHz, $J_x = 0.065$MHz, $J_y = 0.007$MHz and $J_z = 0.004$MHz and hence $|B_{\text{tot}}| > |J_x|$, whereas figure \ref{run2}{\bf b} shows $p(a_1)$ and $p( \downarrow_1 )$ for an effective Hamiltonian (\ref{XYZ}) with the same $J_x$, $J_y$ and $J_z$ but $B_{\text{tot}} = -0.025$MHz
and hence  $|B_{\text{tot}}| < |J_x|$ \cite{HRP06}.

Discrepancies between numerical results for the full and the effective model are due to higher order terms for the parameters $B$, $\tilde{B}$, $J_x$, $J_y$ and $J_z$, which lead to relative corrections of up to 10\% in the considered cases. Despite this lack of accuracy of the second order approximations, the effective model is indeed a spin-1/2 Hamiltonian as occupations of excited atomic and photon states are negligible.

\subsubsection{Experimental requirements} 

For an experimental implementation, the parameters of the effective Hamiltonian, $J_x$, $J_y$, $J_z$, $B$ and $\tilde{B}$ have to be much larger than rates for decay mechanisms via the photons or the excited states $\ket{e_j}$. With the definitions $\Omega = \text{max}(\Omega_a, \Omega_b)$, $g = \text{max}(g_a, g_b)$, $\Delta = \text{min}(\Delta_a, \Delta_b)$, the occupation of the excited levels $\ket{e_j}$ and the photon number $n_p$ can be estimated to be $\langle \proj{e_j} \rangle \approx |\Omega / 2 \Delta |^2$ and $n_p \approx |(\Omega g / 2 \Delta) \gamma_1 |^2$, whereas the couplings $J_x$, $J_y$ and $J_z$ are approximately $|(\Omega g / 2 \Delta)|^2 \gamma_2$.

Spontaneous emission from the levels $\ket{e_j}$ at a rate $\Gamma_E$ and cavity decay of photons at a rate $\Gamma_C$ thus lead to effective decay rates
\begin{equation*}
\Gamma_1 = |\Omega / 2 \Delta |^2 \Gamma_E \hspace{0.15 cm} \text{and} \hspace{0.15 cm} \Gamma_2 = |(\Omega g / 2 \Delta) \gamma_1 |^2 \Gamma_C.
\end{equation*}
The effective dynamics of Hamiltonian (\ref{XYZ}) can thus be observed if 
\begin{equation*}
\Gamma_1 \ll |(\Omega g / 2 \Delta)|^2 \gamma_2  \hspace{0.15 cm} \text{and} \hspace{0.15 cm} \Gamma_2 \ll |(\Omega g / 2 \Delta)|^2 \gamma_2 ,
\end{equation*}
which implies $\Gamma_E \ll J_C \, g^2 / \delta^2$ and $\Gamma_C \ll J_C$
($J_C < \delta/2$), where, $\delta =  |(\omega_a + \omega_b)/2 - \omega_C|$ for the XX and YY interactions and $\delta =  |\omega - \omega_C|$ for the ZZ interactions and the approximations $|\gamma_1| \approx \delta^{-1}$ and $|\gamma_2| \approx J_C \delta^{-2}$ have been used.
Since photons should be more likely to tunnel to the next cavity than decay into free space,
$\Gamma_C \ll J_C$ should hold in most cases. For $\Gamma_E \ll J_C g^2 / \delta^2$,
to hold, cavities with a high ratio $g / \Gamma_E$ are favourable. Since $\delta > 2 J_C$, the two requirements together imply that the cavities should have a high cooperativity factor $\xi$.

\subsection{Magnetic dynamics of polarised light}

In contrast to identifying the internal states of the atoms with spin polarisations,
magnetic dynamics can also be studied by considering cavities which support two distinct
modes of light with different polarisations, $\uparrow$ and $\downarrow$.
In \cite{JXL07} the dynamics of the photon polarisations in such a scheme was analysed.
The authors consider cavities that are doped with atoms of a $V$ level structure, where two distinct
internal transitions only couple to photons of different polarisations.
The interactions of the photons with the atoms are chosen such that double occupancies
are suppressed by the blockade effects discussed in section \ref{sec:JC} and the dynamics of the photons with creation operators $a_{\uparrow}^{\dagger}$ and $a_{\downarrow}^{\dagger}$ for the two polarisations obeys the effective Hamiltonian \cite{JXL07}
\begin{equation} \label{ham_JXL07}
H = \sum_{<j, j'>} J (S_j^x S_{j'}^x + S_j^y S_{j'}^y ) + J' S_j^z S_{j'}^z ,
\end{equation}
where $S_j^z = \frac{1}{2} (a_{j\uparrow}^{\dagger} a_{j\uparrow} - a_{j\downarrow}^{\dagger} a_{j\downarrow})$,
$S_j^x = \frac{1}{2} (a_{j\uparrow}^{\dagger} a_{j\downarrow} + a_{j\downarrow}^{\dagger} a_{j\uparrow})$,
$S_j^y = \frac{1}{2 \iu} (a_{j\uparrow}^{\dagger} a_{j\downarrow} - a_{j\downarrow}^{\dagger} a_{j\uparrow})$ and the indices $j$ and $j'$ label the cavities.

\subsection{Cluster state generation} \label{sec:cluster}

Effective spin models in arrays of coupled cavities can be used for the generation of cluster states \cite{RB01}, which form, together with the local addressability, a platform for one-way quantum computation \cite{RB01a,GE06}. The experimental verification that the desired entanglement has indeed been created is thereby a non-trivial task in its own right \cite{AP06}.

One way to generate cluster states is via the effective Hamiltonian (\ref{HZZ}) \cite{HBP07a}. A similar rout has been presented in \cite{LGG07}.
To this end, all atoms are initialised in the states $(\ket{a_j} + \ket{b_j})/\sqrt{2}$,
which can be done via a STIRAP process \cite{FIM05}, and then evolved 
under the Hamiltonian (\ref{HZZ}) for $t = \pi / 4 J_z$.

\begin{figure}
\centering
\psfrag{A}{\bf \hspace{-2.1cm} a}
\psfrag{B}{\bf \hspace{-2.1cm} b}
\psfrag{t}{\raisebox{-0.4cm}{\scriptsize \hspace{-0.6cm} $t \: \text{in} \: 10^{-6} \: \text{s}$}}
\psfrag{pu}{\raisebox{0.5cm}{\scriptsize \hspace{-0.2cm} $P_{\text{s}}$}}
\psfrag{vN}{\raisebox{0.4cm}{\scriptsize \hspace{-0.6cm} $E_{\text{vN}} / \ln 2$}}
\psfrag{0a}{\raisebox{-0.12cm}{\scriptsize $0$}}
\psfrag{10a}{\raisebox{-0.12cm}{\scriptsize $10$}}
\psfrag{20a}{\raisebox{-0.12cm}{\scriptsize $20$}}
\psfrag{30a}{\raisebox{-0.12cm}{\scriptsize $30$}}
\psfrag{40a}{\raisebox{-0.12cm}{\scriptsize $40$}}
\psfrag{50a}{\raisebox{-0.12cm}{\scriptsize $50$}}
\psfrag{60a}{\raisebox{-0.12cm}{\scriptsize $60$}}
\psfrag{0}{\hspace{-0.4cm} \scriptsize $0$}
\psfrag{0.2}{\hspace{-0.4cm} \scriptsize $0.2$}
\psfrag{0.4}{\hspace{-0.4cm} \scriptsize $0.4$}
\psfrag{0.6}{\hspace{-0.4cm} \scriptsize $0.6$}
\psfrag{0.8}{\hspace{-0.4cm} \scriptsize $0.8$}
\psfrag{1}{\hspace{-0.24cm} \scriptsize $1$}
\psfrag{0.996}{\hspace{-0.55cm} \scriptsize $0.996$}
\psfrag{0.997}{\hspace{-0.55cm} \scriptsize $0.997$}
\psfrag{0.998}{\hspace{-0.55cm} \scriptsize $0.998$}
\psfrag{0.999}{\hspace{-0.55cm} \scriptsize $0.999$}
\includegraphics[width=.46\linewidth]{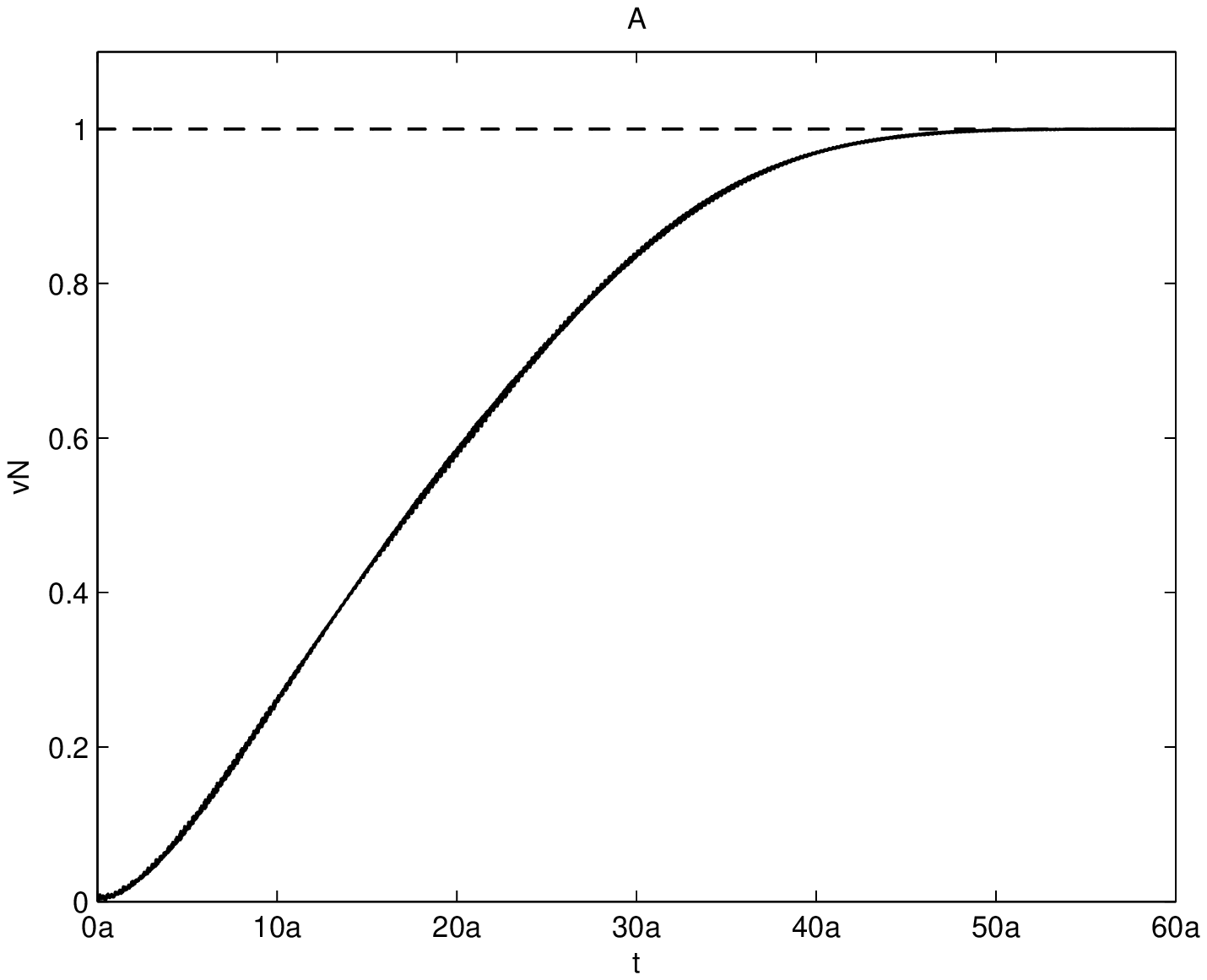}
\hfill
\includegraphics[width=.46\linewidth]{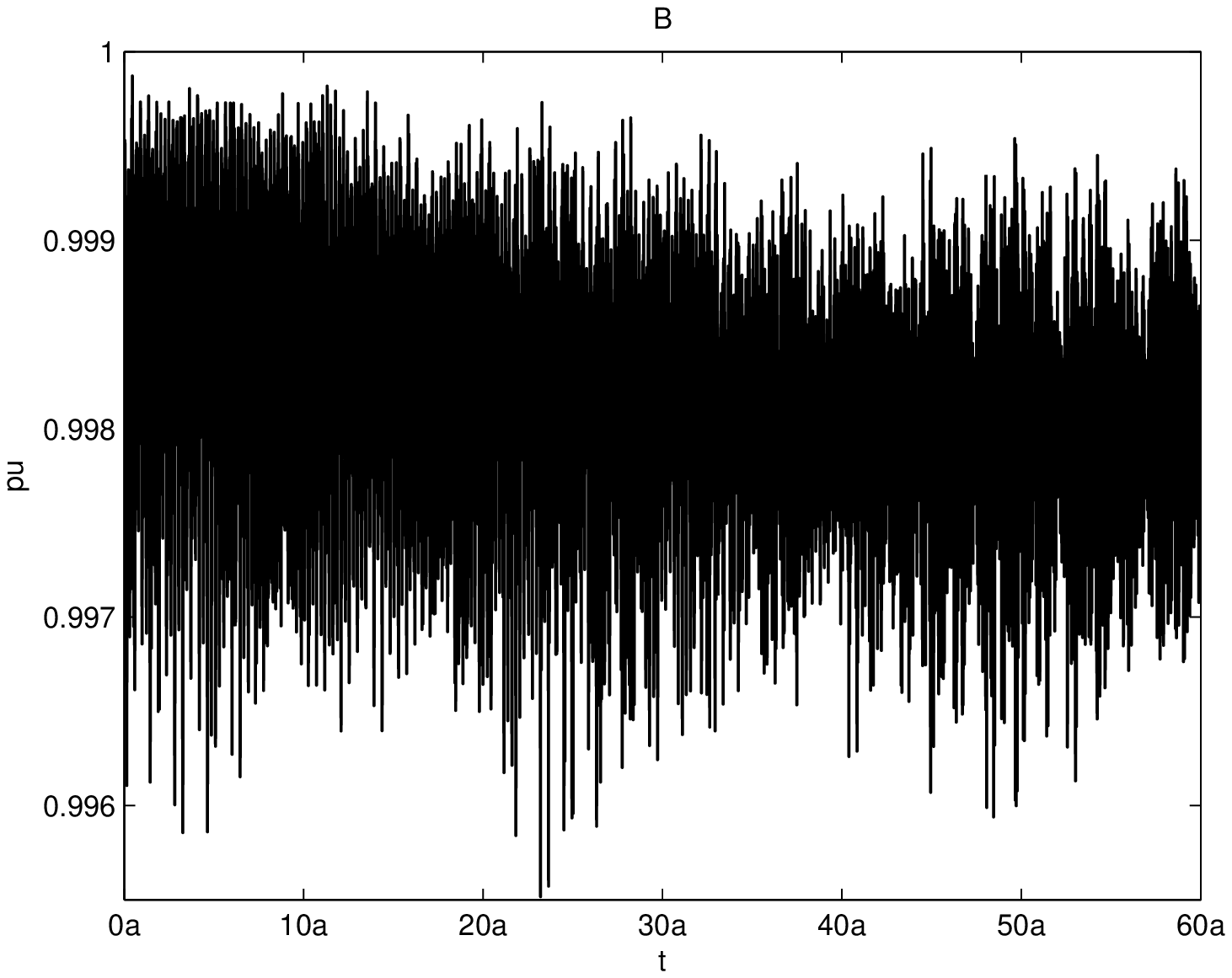}
\caption{{\bf a}: The von Neumann entropy $E_{\text{vN}}$ of the reduced density matrix of 1 effective spin in multiples of $\ln 2$ and {\bf b}: the purity of the reduced state of the effective spin model for 3 cavities where $J_z = 0.021$MHz. The plots assume that no spontaneous emission took place.
For a spontaneous emission rate of $0.1$MHz ($g = 1$GHz), the probability for a decay event in the total time range is 1.5\%. Hence, cluster state generation fails with probability $0.005 \times n$ for $n \ll 1/0.005 = 200$ cavities, irrespectively of the lattice dimension.
Reprinted with permission from \cite{HBP07a}.}
\label{cluster}
\end{figure}

Figure \ref{cluster} shows the von Neumann entropy of the 
reduced density matrix of one effective spin $E_{\text{vN}}$ 
and the purity of the reduced density matrix of the effective 
spin chain $P_{\text{s}}$ for a full three cavity model. Since 
$E_{\text{vN}} \approx log_2 2$ for $t \approx 50 \mu$s while 
the state of the effective spin model remains highly pure 
($P_{\text{s}} = tr[\rho^2] > 0.995$) the degree of entanglement
will be very close to maximal. Thus the levels  $\ket{a_j}$ and $\ket{b_j}$
have been driven into a state which is, up to local unitary rotations, 
very close to a three-qubit cluster state.

\begin{figure*}
 \sidecaption
 \includegraphics[width=.7\textwidth]{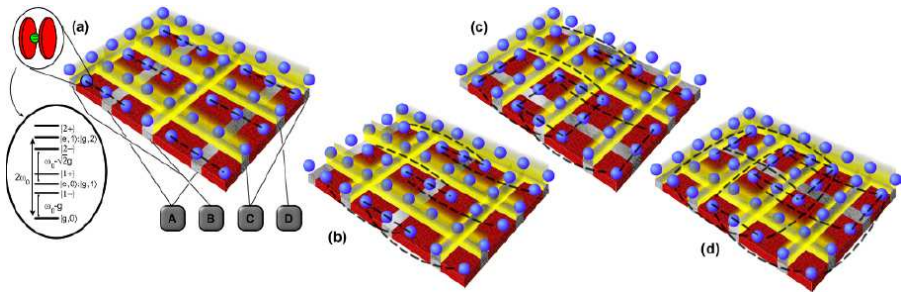}
 \caption{A 2d cavity array. Consecutive application of gates A, B, C and D, depicted in parts (a), (b), (c) and (d), each isolate chains of 3 qubits, realising controlled-phases and SWAPs (swapping quantum/classical gates) between the qubits at either end of the chain (the dashed lines indicate where the controlled-phase gates have been applied).
          Reprinted with permission from \cite{AK08}.}
 \label{cluster2}
\end{figure*}

A second way to generate cluster states is to implement controlled phase gates between
the qubits in different cavities \cite{AK08}. First all qubits are initialised in the state
$\ket{+} = (\ket{1} + \ket{0})/\sqrt{2}$, where $\ket{1}$ and $\ket{0}$ refer to states with
one or zero excitations in a cavity. The phase gates are then implemented by isolating chains
of three qubits via a Stark shift that turns surrounding qubits off resonance.
The isolated three-qubit chains evolve for a time $t = \pi /(2 \sqrt{2} J)$ ($J$ is the photon
tunnelling rate) and the central qubit is subsequently measured. Local unitaries on the remaining
two qubits that depend on the outcome of this measurement complete the controlled phase gate.
Figure \ref{cluster2} shows the entire sequence of gates.

Further ways to create entangled states include measurement based schemes. One version uses two atoms in two cavities that interact via the exchange of photons \cite{CAB07}. Both atoms can emit photons of two polarisations (left circular and right-circular), depending on which excited state they are in. Hence the two atoms are prepared such that they each emit a single photon, one with left-circular and one with right-circular polarisation. After a time $t = \pi / 4 J$ ($J$ being the photon tunnelling rate), when photons have tunnelled between the cavities with probability $1/2$, the two atoms are employed to perform a non-demolition measurement to determine the number of photons in each cavity. Provided each cavity contains one photon the state is collapsed to a two photon polarisation entangled state thus providing a heralded source of entangled photons.

Apart from quantum information motivated applications coupled cavity arrays with local
addressability can also be used to study novel physical setups. Among those are interface between
regions in different condensed matter phases, see section \ref{sec:appl}.


\section{Promising Platforms for Realisations} \label{sec:impl}

In this section, we present some cavity QED setups which have favourable features for the realisation of the effective many-body models described in this review. These platforms are thus promising candidates for experimental demonstrations of the theory presented here. The selection we present here obviously reflects our personal view.
It is of course possible that devices which we do not explicitly mention here may in the end provide the first realisation of polaritonic many-body systems as discussed in previous sections.

A major part of all cavity QED experiments have been performed with Fabry-P\'erot cavities formed by rather massive mirrors. With such mirrors, coupling two cavities such that photons can tunnel between them seems quite difficult. On the other hand several new cavity QED structures with small volume micro-cavities have emerged in recent years. These are routinely produced in large arrays and by construction can allow for efficient photon tunnelling between neighbouring cavities. In addition, the strong coupling regime with cooperativity factors much larger than unity has been achieved in several of these devices. These two features together make such micro-cavity arrays very promising candidates for the generation of the effective many-body models discussed in this review. In the following we discuss some platforms in a bit more detail.

\subsection{Photonic band gap cavities}

One promising device consists of quantum dots coupled to photonic band gap defect nanocavities. Photonic crystals are structures with periodic dielectric properties which affect the motion of photons, in a similar manner as the periodicity of a semiconductor affects the motion of electrons \cite{JJ02}. Such structures can exhibit band gaps in frequency space in which no photon of particular frequencies can propagate in the material (for an introductory review see \cite{AKcontph}).

\begin{figure}
 \sidecaption
 \includegraphics[width=.5\linewidth]{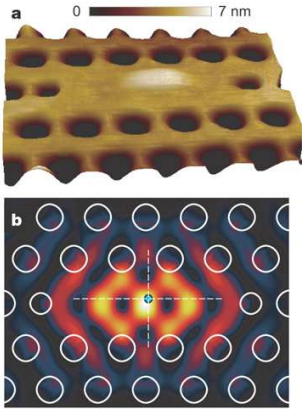}
 \caption{A quantum dot in a photonic band gap cavity. {\bf a}, AFM topography of a photonic crystal nanocavity aligned to a hill of material on the surface arising from a quantum dot buried 63 nm below. {\bf b}, Electric field intensity of the photonic crystal cavity mode showing that the location of the buried QD, indicated by the teal dot, overlaps the field maximum. Reprinted with permission from \cite{Hen07}.}
 \label{pbg_Hen07}
\end{figure}

A nanocavity can be created in a photonic-band-gap material by producing a localised defect in the structure of the crystal, in such a way that light of a particular frequency cannot propagate outside the defect area. Large arrays of such nanocavities have been produced \cite{AV04}. Photon hopping between neighbouring cavities has been observed in the microwave and optical domains \cite{BTO00a,BTO00b}, where its description in the form as given in equation (\ref{arrayham2}) was found to be accurate. Quantum dots, semiconductors whose excitations are confined in all three directions, behave in many respects like atoms and can be effectively addressed by lasers \cite{JA98}. Quantum dots can be created inside photonic crystal nanocavities \cite{Bad05} and made to interact with the cavity mode to form a standard cavity QED system. The strong coupling regime with cooperativity factors up to $\xi \approx 10$ has already been achieved in such systems \cite{Hen07}. Cooperativity factors of $\xi \sim 10^6$ are predicted to be realisable \cite{Spil05}. Due to the extremely small volume of the nanocavity, the Jaynes-Cummings coupling coefficient can be extremely large \cite{AAS+03,SNAA05}, while spontaneous emission rates are very low. Figure \ref{pbg_Hen07} shows a quantum dot in a photonic band gap cavity.
For the applications discussed in this review current cavity decay rates of photonic band gap cavities remain a limiting factor. This is mainly due to the fact that most structures employ two dimensional photonic crystals as placing a quantum dot into a three dimensional structure is difficult. The difficulties to control the properties of individual quantum dots and to place them at desired locations could even pose a more significant problem. Current experiments pick a quantum dot from an ensemble and
then fabricate a photonic crystal tailored to this dot around it. Such an approach can obviously not be used for producing a whole array of sufficient quality.

One might think about trapping atoms in photonic band gap cavities instead of letting them interact with quantum dots. Since the frequencies of currently available photonic band gaps do not fit to atomic transition frequencies, such a strategy would require the development of new photonic band gap materials.

\subsection{Silicon structures: micro discs and micro toroids}

\begin{figure}
 \sidecaption
 \includegraphics[width=.5\linewidth]{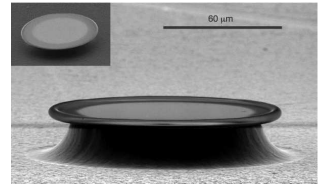}
 \caption{SEM picture of a toroidal microresonator with an intrinsic cavity Q of $10^8$. Reprinted with permission from \cite{AKS+03}.}
 \label{microtoroid_AKS+03}
\end{figure}

A second class of micro-cavities that are routinely produced in large arrays and which
have very high Q-factors are silicon structures of either a disc \cite{BSP+06} or a toroidal \cite{AKS+03} shape, see figure \ref{microtoroid_AKS+03}. These cavities are closed silicon discs or toroids that typically are carried by a thin pillar in their centre. They are routinely produced in large arrays \cite{ASV07}. The light is trapped in whispering gallery modes that are localised close to the outer surface of the structure and have a small mode volume \cite{KSMV04} . Very efficient and highly tunable coupling of photons into and from those cavities is possible via tapered optical fibres that are placed close to their surface \cite{SKPV03}. Here the evanescent field of both cavity and fibre overlap and photons can hop between both. The photon hopping rate can thereby be controlled very accurately via the separation between cavity and fibre.

These cavities can interact with atoms in a strong coupling regime with cooperativity factors $\xi \sim 50$ \cite{ADW+06} if the atoms are placed close to the cavity surface and interact with the evanescent field. This has been seen in an experiment where atoms fell by the cavity surface \cite{ADW+06} and a photon blockade effect generated by one atom has been demonstrated \cite{Day08}. Theoretical predictions \cite{Spil05} show that very large cooperativity factors of $\xi \sim 10^6$ are expected to be achieved in such systems.

For realising effective Hamiltonians such as (\ref{bosehubbard}), (\ref{JC_eff_mb}) or (\ref{XYZ}), some aspects still remain challenging for micro disc and micro toroidal cavities. One aspect is that atoms need to be trapped for a sufficiently long time close enough to the cavity surface to maintain a sufficient strong coupling regime.
A second aspect is that all cavities of an array need to be tuned into resonance with each other with enough accuracy. This can be achieved by varying their temperature and thus
shrinking or expanding them, but is still a technical challenge. Finally there can be modifications of the photon hopping as compared to the form assumed in equation (\ref{arrayham2}). A challenge here would be to prevent that significant photon population is localised in the fibre instead of in the cavities themselves.
Nonetheless, even if the fibre contains appreciable excitations, interesting many-body physics is expected to happen as the fibre would merely play the role of an additional lattice site with different local terms than the other sites.

\begin{figure}
 \centering
 \includegraphics[width=.9\linewidth]{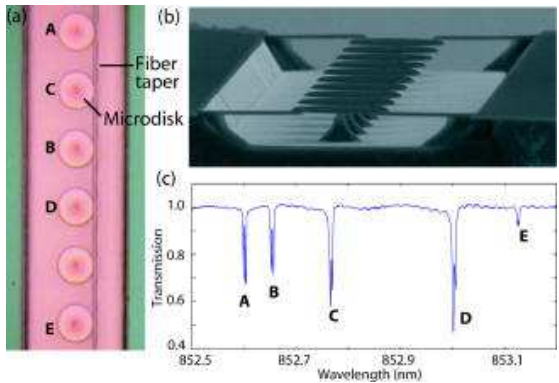}
 \caption{(a) Top-view optical image of a fibre taper aligned with and array of ten micro-disks (the remaining four micro-disks are out of the field of view of the image). (b) SEM image of the array of ten micro-disks. (c) Fibre taper transmission vs wavelength when the taper is aligned with the micro-disk array in (a). The letters match specific micro-disks in (a) with the corresponding resonances in (c). Reprinted with permission from \cite{BSP+06}.}
 \label{discarray_BSP+06}
\end{figure}

Figure \ref{discarray_BSP+06} shows an array of 10 micro disc cavities that are all coupled by one tapered optical fibre and mounted on an atom chip. The purpose of the atom chip is to trap atoms close to the cavity surfaces and, in that way, maintain a strong coupling regime for substantial times.

\subsection{Fibre based cavities}
\begin{figure}
 \centering
 \includegraphics[width=\linewidth]{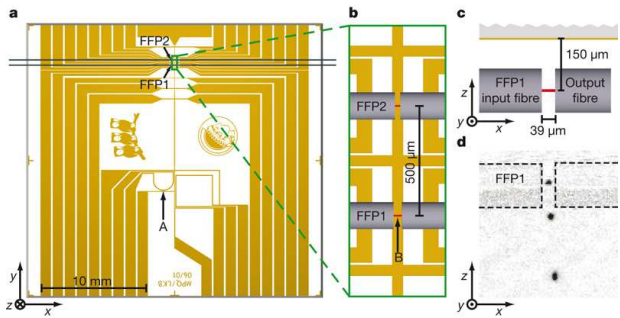}
 \caption{{\bf a}, Layout of an atom chip that contains two fibre Fabry–P\'erot cavities. {\bf b}, Close-up view of the two fibre Fabry–P\'erot (FFP) optical cavities that are mounted on the chip. Cavity modes are drawn to scale in red. The BEC is produced in a magnetic trap and positioned in the FFP1 mode ('B'). c, Geometry of the FFP1 cavity. d, Overlay of three CCD time-of-flight (TOF) absorption images, showing the anisotropic expansion of a BEC having interacted for 50 ms with the cavity field. The optical fibres are outlined for clarity. Reprinted with permission from \cite{CSD+07}.}
 \label{fibrefp_CSD+07}
\end{figure}
Strong coupling experiments with many atoms in one cavity and at the same time large cooperativity factors for each individual atom have been performed with fibre Fabry-P\'erot cavities \cite{CSD+07}. These cavities are formed by two tips of optical fibres that are facing each other and are mounted
on an atom chip to trap an manipulate the atoms that interact with the cavity modes.
Current chips contain two such cavities as shown in figure \ref{fibrefp_CSD+07} and cooperativity factors as high as $\xi = 145$ have been achieved \cite{CSD+07}.

Photons are coupled into and from the cavity directly via the optical fibres.
Hence as for fibre coupled micro discs or toroids this may lead to modifications of the
form of the photon hopping as compared to equation (\ref{arrayham2}).
On the other hand, the atoms are here trapped on the atom chip and can thus be held in a location of strong interaction with the cavity modes for a long time.

\subsection{On-chip Fabry-P\'erot cavities}
\begin{figure}
 \centering
 \includegraphics[width=.3\linewidth]{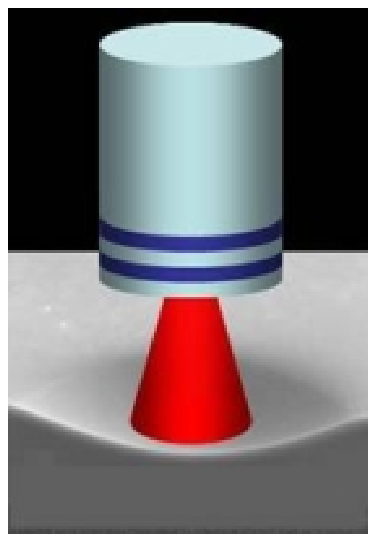}
 \hspace{.1\linewidth}
 \includegraphics[width=.5\linewidth]{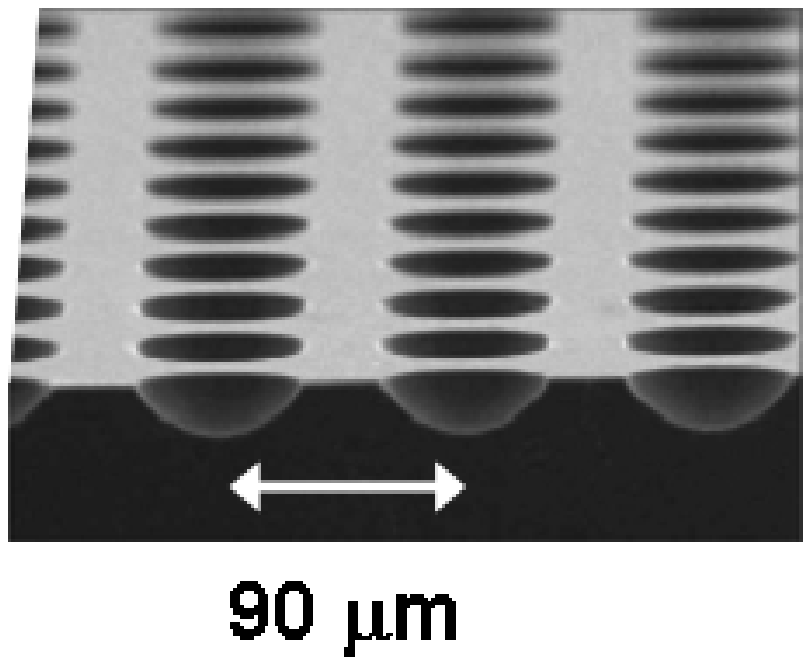}
 \caption{Left: Single on-chip Fabry-P\'erot cavity formed by an etched micro-mirror and a tip of an optical fibre. Right: Two dimensional array of cavity micro-mirrors. Reprinted with permission from \cite{CCM_web}.}
 \label{hinds_CCM_web}
\end{figure}
Micro-fabricated arrays of Fabry-P\'erot type cavities with large cooperativity factors have been presented in \cite{THE+05}. Each plano-concave resonator consists of an etched spherical mirror and a coated single-mode optical fibre, allowing open access to the intra-cavity field mode. Here, large regular arrays of spherical mirrors can be created with controllable sizes and spacings.
These cavities which have a very small mode volume are illustrated in figure \ref{hinds_CCM_web}.
The extremely small mode volume leads to a large single-photon Rabi frequency, and light is conveniently coupled into and out of the cavities via the fibres themselves. These cavities to date achieve a finesse $F \sim 5000$, corresponding to Purcell factors (or cooperativities) $\xi \sim 40$.

As with the other cavities that involve optical fibres modifications of the photon hopping as compared to equation (\ref{arrayham2}) are expected here as well.
Current strong coupling experiments have also been only performed with falling atoms
\cite{TGD+07}.

\subsection{Superconducting stripline resonators}
\begin{figure}
 \centering
 \includegraphics[width=.9\linewidth]{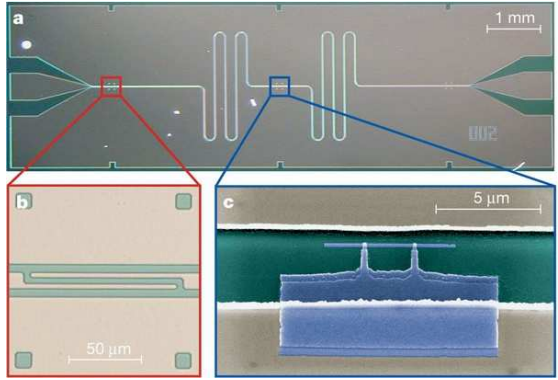}
 \caption{{\bf a}, A superconducting niobium coplanar waveguide resonator of 24 mm length. It is coupled by a capacitor at each end of the resonator (see b) to an input and output feed line. {\bf b}, The capacitive coupling to the input and output lines. {\bf c}, False colour electron micrograph of a Cooper pair box (blue) fabricated onto the silicon substrate (green) into the gap between the centre conductor (top) and the ground plane (bottom) of a resonator (beige) using electron beam lithography and double angle evaporation of aluminium. Reprinted with permission from \cite{Wal04}.}
 \label{stripline_Wal04}
\end{figure}
A promising experimental platform that operates in the microwave regime are cavity QED system formed by a Cooper-pair box coupled to a superconducting transmission line resonator. A Cooper-pair box is formed by two superconducting islands separated by a Josephson junction. Although it is a macroscopic device, it behaves in many ways like an atom as the relevant structure can be described by a two level quantum system. A superconducting transmission line cavity, on the other hand, is a quasi-one-dimensional coplanar waveguide resonator, formed on a chip with a resonance frequency in the microwave range. The strong-coupling regime in such a system has been achieved \cite{Wal04} and recently two Cooper-pair boxes have been strongly coupled to the cavity mode \cite{SPS07,Maj07}. Very large cooperativity factors have been achieved in this set-up \cite{Wal04,Maj07} and the nonlinearity of the Jaynes-Cummings model has been measured \cite{FGB+08}. Although coupling between different cavities has not been realised, up to ten Cooper-pair boxes interacting in the strong coupling regime with the resonant cavity mode has been predicted to be achievable using the architecture of Ref. \cite{Maj07}.


\section{Applications} \label{sec:appl}

Structures of coupled arrays of micro-cavities have several interesting
potential applications. Particularly useful in this respect is the individual
addressability of single cavities with optical lasers. The feasibility of single
site addressing is based on the size of the cavities which can easily be larger
than an optical wavelength. Furthermore the inter-cavity coupling could be realised
by optical fibres which would move the cavities even further apart.

Among the potential applications are the implementation of tools for
quantum information processing. These include the generation of entangled states
such as cluster states \cite{HBP07a,AK08,AMB07,CAB07} (see section \ref{sec:cluster}),
the controlled propagation of polaritonic excitations in cavity arrays
\cite{BAB07,ZLS07,HZSS07,ZGSS08,HLSS08}, quantum gates \cite{ASYE07}
and the generation of single photons
in parallel \cite{NY08}.

\begin{figure}
\centering
\includegraphics[width=.8\linewidth]{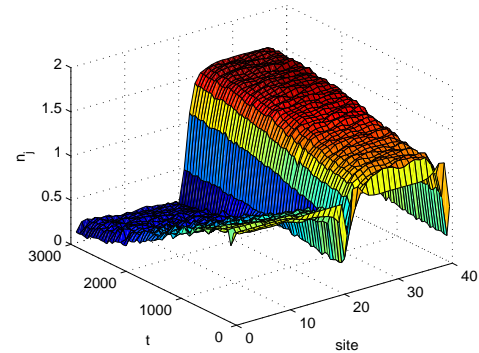}
\caption{The particle density across the chain as function of time.
Sites 1 to 20 start in a Mott insulator regime and the
rest in a superfluid regime. $U = 1.0$, $J = 0.1$, $\mu = 0$, $\tilde{U} = 0.2$, $\tilde{\mu} = 0$, $\tilde{J} = 1.0$ and $J_I = 0.1$. Reprinted with permission from \cite{HP08}.}
\label{numdens1}
\end{figure}

Perhaps most importantly, the local controllability offered by coupled cavity arrays can also give rise to novel physical effects and we will mention here one example of these.
Local manipulations of an effective many body system could be applied to engineer a boundary between a Mott insulator and a superfluid 
region of a Bose-Hubbard model at unit filling. Initially both 
regions are decoupled and cooled to their respective ground
states. This situation can for example be prepared in a cavity array that is operated
in the way as discussed in section \ref{BHmodel} and thus hosts a polaritonic
Bose-Hubbard Hamiltonian (\ref{bosehubbard}). After switching on a small tunnelling rate 
between both regions, all particles of the Mott region migrate 
to the superfluid area. This behaviour is shown for a one dimensional chain in
figure \ref{numdens1}, where the particle densities for all
lattice sites are shown as a function of time. The migration takes place whenever the 
difference between the chemical potentials of both regions is 
less than the maximal energy of any eigenmode of the superfluid \cite{HP08}. 

\section{Summary} \label{conc}

In this review, we have tried to give an overview of the current status of approaches to engineer effective many body dynamics in arrays of cavities that are coupled via photon tunnelling.
The schemes that have been developed so far can be divided into two main categories, effective Bose-Hubbard type models and effective spin models.

In Bose-Hubbard type models, quantum phase transitions between superfluid phases of polaritons or photons (coherent light) and Mott insulators can be studied. Here, the experimentally challenging and hence physically new and interesting part is to create a Mott insulator in which polaritons or photons are by their effective mutual repulsion forced to stay in one lattice site (cavity) and not able to move. This situation would in correspond to a crystal formed by light or ''frozen light''.

A key issue on the experimental side is to generate a Kerr nonlinearity or effective interaction between polaritons/photons that is significantly larger than the decay mechanisms present in the cavity array.

Existing approaches for effective spin models mainly employ two metastable levels of an atom trapped in each cavity to represent the two spin polarisations $\uparrow$ and $\downarrow$.
Raman transitions mediated by external driving lasers that give rise to interactions between neighbouring cavities via the exchange of virtual photons are then used to generate effective spin-spin interactions.

As for Bose Hubbard type models a main experimental challenge is to make atom-photon interactions
significantly stronger than the decay mechanisms in the cavity array.

Effective many body system in a coupled cavity arrays are of great interest from various points of view. First of all they open up a way to study condensed matter physics with light and hence to engineer new quantum states of light. Furthermore they offer a great versatility for applications as quantum simulators due to the possibilities to manipulate them via external driving lasers. 
Most importantly they would, by construction, allow for control, manipulations and measurements of individual lattice sites. The situation is somehow complementary to cold atoms in optical lattices, where very large lattices can be created and experiments provide good measurement access to global quantities such as the phase coherence of the superfluid.

A polaritonic many body system in a cavity array on the other hand will for the foreseeable future only comprise much smaller lattices, but provide good access for the measurement of local quantities. In contrast to optical lattices it would thus allow to directly verify the integer occupation number of each lattice site in the Mott insulator regime.

The local addressability of coupled cavity arrays gives rise to a number of novel application.
These contain quantum information processing application but also novel physical situations that give rise to features such as particle migration.

\appendix

\begin{acknowledgement}
This work is part of the QIP-IRC supported by EPSRC (GR/S82176/0), the Integrated Project Qubit Applications (QAP) supported by the IST directorate as Contract Number 015848', the EU STREP project HIP and was supported by the EPSRC grant EP/E058256/1, the Alexander von Humboldt Foundation, the Conselho Nacional de Desenvolvimento Cient\'ifico e Tecnol\'ogico (CNPq), the Royal Society and the DFG Emmy Noether grant HA 5593/1-1.
\end{acknowledgement}

\end{document}